\PassOptionsToPackage{table}{xcolor}
\documentclass{aa}

\usepackage{colortbl}
\usepackage{graphicx} 
\usepackage{placeins}
\usepackage{booktabs}
\usepackage[table]{xcolor}

\definecolor{lightgray}{rgb}{0.83, 0.83, 0.83}
\definecolor{lightblue}{rgb}{0.67, 0.84, 0.90}
\definecolor{lightgreen}{rgb}{0.56, 0.93, 0.56}

\usepackage{natbib}
\bibpunct{(}{)}{;}{a}{}{,} 

\usepackage{threeparttable}

\usepackage[varg]{txfonts}
\usepackage{hyperref}
\usepackage{multirow}
\usepackage{color}
\definecolor{green}{rgb}{0.3,0.7,0.}
\definecolor{purple}{rgb}{0.77, 0.29, 0.55}

\newcommand{\Ms}{{\ensuremath{M_{\odot} }}}
\newcommand{\Rs}{{\ensuremath{R_{\odot} }}}

\hypersetup{
    colorlinks=true,       
    linkcolor=blue,          
    citecolor=blue,        
    filecolor=blue,      
    urlcolor=blue           
}

\begin{document}

\title{Rotating supermassive Pop III stars on the main sequence}

\titlerunning{rotating supermassive Pop III stars}
\author{Devesh Nandal \inst{1}, Gaël Buldgen\inst{2}, Daniel J. Whalen \inst{3}, John Regan \inst{4}, Tyrone E. Woods \inst{5}, Jonathan C. Tan
          \inst{1,}\inst{6}
          }
\authorrunning{Nandal et al}

\institute{Department of Astronomy, University of Virginia, 530 McCormick Rd, Charlottesville, VA 22904, USA \and STAR Institute, Universit{\'e} de Li{\`e}ge, Li{\`e}ge, Belgium \and Institute of Cosmology and Gravitation, Portsmouth University, Dennis Sciama Building, Portsmouth PO1 3FX \and Centre for Astrophysics and Space Sciences Maynooth, Department of Physics, Maynooth University, Maynooth, Ireland \and Department of Physics \& Astronomy, Allen Building, 30A Sifton Rd,
University of Manitoba, Winnipeg MB  R3T 2N2, Canada \and Dept. of Space, Earth and Environment, Chalmers Univ. of Technology, Gothenburg, Sweden}

\date{}

\abstract{ The detection of billion-solar-mass supermassive black holes (SMBHs) within the first billion years of cosmic history challenges conventional theories of black hole formation and growth. Simultaneously, recent JWST observations revealing exceptionally high nitrogen-to-oxygen abundance ratios in galaxies at high redshifts raise critical questions about rapid chemical enrichment mechanisms operating in the early universe. Supermassive stars (SMSs) with masses of 1000 to 10000 \Ms\ are promising candidates to explain these phenomena, but existing models have so far neglected the pivotal role of stellar rotation. Here, we present the first comprehensive evolutionary models of rotating Pop III SMSs computed using the GENEC stellar evolution code, including detailed treatments of rotation-induced chemical mixing, angular momentum transport, and mass loss driven by the $\Omega\Gamma$ limit. We demonstrate that rotation significantly enlarges the convective core and extends stellar lifetimes by up to 20\%, with moderate enhancement of mass-loss rates as stars approach critical rotation thresholds. Our results further indicate that the cores of SMSs rotate relatively slowly (below $\sim 200$ km s$^{-1}$), resulting in dimensionless spin parameters $a* < 0.1$ for intermediate-mass black hole (IMBH) remnants that are notably lower than theoretical maximum spins. These findings highlight rotation as a key factor in determining the structural evolution, chemical yields, and black hole spin properties of SMSs, providing critical insights to interpret observational signatures from the high-redshift universe.
}

\keywords{Stars: evolution -- Stars: Population III -- Stars: massive -- Stars: rotation}

\maketitle

\section{Introduction}\label{Sec:Introduction}

The discovery of luminous quasars hosting billion-solar-mass supermassive black holes (SMBHs) at redshifts $z \gtrsim 7$ poses significant theoretical challenges to our understanding of black hole formation and growth in the early universe \citep{Mortlock2011, Wu2015, Banados2018}. The existence of these massive objects within the first billion years necessitates rapid formation pathways, motivating models invoking heavy seed black holes originating from direct collapse of primordial gas into supermassive stars (SMSs) in the mass range of $10^3$–$10^5$ \Ms\ \citep{Latif2016,Smidt18,Inayoshi2020,Latif_2022,pat23a}. Complementary evidence from recent JWST observations further complicates this picture, revealing galaxies at $z > 8$ with anomalously high nitrogen-to-oxygen ratios (N/O), significantly exceeding standard enrichment scenarios \citep{Cameron2023, Nakajima2023, Isobe2023, Larson2023}. These observations suggest the early universe hosted exotic stellar populations capable of rapid chemical enrichment \citep{Bunker2023, Marques2023, Liu2025}.

Several mechanisms have been proposed to explain these chemical anomalies, including enrichment by Wolf–Rayet starbursts, asymptotic giant branch (AGB) stars, and extremely massive stars \citep{Vink2023, Nandal2024}. Wolf–Rayet stars and AGB stars, while prolific nitrogen producers, typically require metallicities or evolutionary timescales that are incompatible with observed high N/O ratios at such early epochs \citep{Cameron2023}. Alternatively, SMSs are compelling candidates due to their potential to synthesize and rapidly disperse large amounts of primary nitrogen, produced through efficient CNO cycling in their cores, into their surrounding environments via strong mass-loss events or explosive endpoints \citep{Nagele2023, Nandal2025}.

Indeed, hydrodynamical simulations have demonstrated viable conditions for SMS formation possibly via strong external Lyman-Werner radiation \citep{Latif_2014b,Schauer_2017, Regan_2017}, dynamically heated environments \citep{Regan_2017, Wise2019, Regan_2020b}, or through the suppression of fragmentation by baryon-dark matter streaming motions \citep{Schauer2017,Liu2024, Ishiyama2025}. More recent simulations also show that rapid mass growth via cold accretion flows can lead to SMS formation \citep{Latif2022, Kiyuna2023, Kiyuna2024}. Such conditions permit the formation of very massive objects and subsequently massive seed black holes, directly addressing the SMBH timing problem \citep{Woods2017, Whalen2020}. In instances where the host dark matter halo can grow to sufficient masses, the gas inflow rate into the central regions of the halo can exceed 10$^{-2}$-10$^{-1}$ \Ms\ yr$^{-1}$, allowing SMSs to form in such halos. 
An alternative pathway that facilitates such high accretion rates is the Population III.1 (Pop III.1) model, which invokes heating from Weakly Interacting Massive Particle (WIMP) annihilation \citep{Spolyar2008,Tan2024}. In this scenario, the slow collapse of gas within an isolated, pristine minihalo can adiabatically contract the co-located dark matter, boosting its density to a level where WIMP annihilation becomes a dominant energy source for the forming protostar \citep{Banik2019}. This energy injection supports the protostar in an inflated, "swollen" state, maintaining a cool photosphere ($\sim 10^4$ K) that emits few ionizing photons \citep{Nandal2025c}. The resulting suppression of radiative feedback allows the star to bypass the mass limit typical of standard Pop III stars and grow efficiently, potentially reaching supermassive scales of $\sim 10^5$ \Ms\ by accreting the bulk of its host minihalo's baryonic content \citep{Singh2023}.

In cases \citep[where high accretion rates can be maintained;][]{latif21a}, protostellar growth to \(\sim10^5\,M_\odot\) can be achieved before the general-relativistic instability triggers collapse to a massive black-hole seed \citep{Banik2019, Wise_2019, Regan_2020, Regan_2022, Singh2023,herr23a}. However, current stellar evolution models for SMSs assume no rotation and thus neglect centrifugal support and rotationally induced chemical mixing. State-of-the-art stellar evolution models also omit rotationally enhanced mass loss and angular momentum transport that determines black hole spin, precluding theoretical predictions of remnant spin distributions \citep{Woods2020, Nandal2023}.

Rotation significantly influences massive star evolution by enhancing internal mixing through meridional circulation and shear instabilities, enlarging convective cores, and extending stellar lifetimes \citep{Maeder1997, Heger2000, Ekstrom2012, Nandal2024b}. In standard massive stars ($\sim 10$–120 \Ms), rotation can induce chemically homogeneous evolution, drastically altering their evolutionary pathways and final fates \citep{Yoon2005, Yoon2012}. Additionally, rotational mixing affects stellar surface abundances, enriching atmospheres with processed elements such as helium and nitrogen much earlier than in non-rotating stars \citep{Brott2011, Limongi2018, Murphy2021a,Tsiatsiou2024, Nandal2024e}. Rotation also enhances mass-loss rates through mechanical winds, particularly near critical rotation thresholds, thereby significantly influencing stellar mass evolution and the resulting remnant masses \citep{Langer1998, Vink2018}.

Near the Eddington limit, the interplay between radiation pressure and rotation results in the $\Omega\Gamma$ limit, a threshold where effective gravitational acceleration at the stellar surface nearly vanishes \citep{Langer1998, OGlimit2000}. Stars approaching this limit experience drastically enhanced mass loss, even at rotation rates below classical break-up velocities \citep{Glatzel1998, Owocki2004}. This rotational-radiative instability is particularly relevant for SMSs, given their inherently high luminosities near the Eddington limit, potentially imposing strict caps on their rotation and significantly influencing their evolution and ultimate fates \citep{Owocki2004, Sanyal2015}.

Rotation critically impacts the angular momentum content and resulting spins of black hole remnants from massive stars \citep{Heger2000}. Intermediate-mass black holes (IMBHs), including those from SMS collapse, inherit angular momentum profiles established during stellar evolution \citep{Woosley2006, Fryer2011}. Black hole spins strongly affect accretion disk dynamics, jet production, and gravitational-wave emission during mergers, thereby influencing their observational signatures and theoretical interpretations \citep{Berti2008, Barausse2012}. Understanding the rotational evolution of progenitor stars thus provides crucial constraints on IMBH spins, with direct implications for SMBH formation scenarios through hierarchical merging \citep{Sesana2009}.

To date, no models of SMSs above 1000 \Ms\ with rotation exist, leaving substantial uncertainties in predictions of their evolution and ultimate fates. Motivated by observational puzzles and theoretical imperatives, we present in this paper the first comprehensive set of rotating evolutionary models for Population III SMSs with masses of 1000 - 10000 \Ms. Our models incorporate detailed treatments of rotation-induced mixing, angular momentum transport, and the effects of the $\Omega\Gamma$ limit.

In Section~\ref{Sec:Methods} we describe our numerical methods and input physics, including the implementation of rotation in the stellar evolution code GENEC. Section~\ref{Sec:Results} presents detailed evolutionary results, highlighting the influence of rotation on stellar lifetimes, internal structures, and mass loss. In Section~\ref{Sec:Discussion}, we compare our rotating SMS models to previous non-rotating calculations and discuss their implications for early chemical enrichment and SMBH seed formation scenarios. Finally, in Section~\ref{Sec:Conclusion}, we summarize our key findings and outline promising directions for future research.

\section{Methods}\label{Sec:Methods}
We present the first rotating models of supermassive (1000-10000 \Ms) Pop~III stars and investigate the effects of the $\Omega\Gamma$ limit on their evolution. These models use the same numerical setup and physics prescriptions as in prior non-rotating studies \citep{Nandal2024d}, differing only by the inclusion of rotation and thereby isolating its effects. All the models are computed using the Geneva Stellar Evolution Code \citep[GENEC;][]{Eggenberger2008, Ekstrom2012, Nandal2024b}.

\subsection{Initial ZAMS structures}
For stellar masses exceeding about 1000 \Ms, the notion of a classical Zero Age Main Sequence (ZAMS) is not well defined. We therefore begin by constructing non-rotating, accreting pre-MS Pop~III models using the tracks of \citet{Nandal2025}. These tracks begin at 2 \Ms\ and, under constant accretion rates, grow until the target mass (between $1000$ \Ms\ and 10000 \Ms) is attained. While accretion in nature is highly variable \citep{Regan_2020}, the final ZAMS structure is primarily determined by the star's total mass and composition, as it reaches thermal equilibrium during the final contraction phase, largely erasing the memory of its prior accretion history \citep{Nandal2023}. Despite remarkable successes in the 3-D hydrodynamic simulations that account for angular momentum transport in the pre-MS \citep{Hirano2018,Kimura2023}, a self-consistent 1D prescription for stellar evolution codes does not yet exist in the literature. Therefore, our pre-MS models do not include rotation. In addition, this allows us to generate a consistent set of non-rotating ZAMS models to serve as a baseline for our study.

Rotation is then introduced with surface velocities of 0.01, 0.05, and 0.10 of the critical Keplerian velocity ($v/v_{\mathrm{crit}}$). These relatively low rates were chosen specifically to test the regime where the $\Omega\Gamma$ limit is predicted to become important for these massive stars \citep{OGlimit2000, Lionel2021} and to quantify its effect from the onset of the main sequence. The construction of these initial rotating models is detailed below.

The initial structure for each rotating model is generated using a dedicated numerical procedure with three key steps:
\begin{enumerate}
    \item Contraction to the ZAMS: Once a non-rotating pre-MS model reaches its final target mass, accretion is terminated. The model contracts on a Kelvin–Helmholtz timescale until it reaches its highest effective temperature, at which point its structure is in near-thermal equilibrium. The resulting composition is $X=0.7516$; $Y=0.2484$; $Z=0$. The physical parameters of this non-rotating ZAMS model, including the profiles of pressure, temperature, density, and chemical composition are then extracted.
    \item Incorporation of Rotation: The extracted non-rotating structure serves as the input for a new GENEC simulation via a dedicated \textsc{Fortran} routine. At the start of this run, angular momentum is injected to produce the desired surface rotation rate (e.g., $0.1~v/v_{crit}$). The mapping from the non-rotating structure to the rotating one is handled by a dedicated \textsc{Fortran} wrapper that passes the full radial grids of $P$; $T$; $\rho$; and $X_i$ to \textsc{GENEC}. The code then self-consistently evolves the transport of angular momentum and chemical species.
    \item Relaxation to a Rotating Equilibrium: After rotation is introduced, the model evolves for several thousand years. This timescale is negligible compared to the main-sequence lifetime but is sufficient for the star to dynamically and thermally relax to a stable, rotating equilibrium. We define this converged state as the rotating ZAMS. The model settles on the blue side of the HR diagram at its maximum $T_{\mathrm{eff}}$, which we adopt as the ZAMS reference point. This procedure ensures our investigation begins from a physically consistent starting point.
\end{enumerate}
The main sequence is defined to begin when the central hydrogen mass fraction, $X_c$, has decreased by $10^{-3}$ from its initial value, and to end when $X_c$ falls below $10^{-3}$. These self-consistent initial models set the stage for our rotational prescription, described next.

\subsection{Treatment of rotation}
We adopt the shellular rotation framework of \citet{Zahn1992}, in which strong horizontal turbulence enforces nearly uniform angular velocity on isobaric surfaces. The transport of chemical species is described by a purely diffusive equation:
\begin{equation}
\rho \,\frac{\partial X_i}{\partial t}\Bigg|_{M_r} \;=\; \frac{1}{r^2}\,\frac{\partial}{\partial r}\left(\rho \,r^2 \,D_{\rm chem}\,\frac{\partial X_i}{\partial r}\right),
\label{eq:chem_transport}
\end{equation}
where $X_i$ is the mass fraction of species $i$, 
\begin{equation}
D_{\rm chem} \;=\; D_{\rm shear} \;+\; D_{\rm eff},
\end{equation}
and 
\begin{equation}
D_{\rm eff} = \frac{1}{30}\,\frac{\bigl[r\,U_2(r)\bigr]^2}{D_{\rm h}}.
\label{eq:Dh}
\end{equation}
We take $D_{\rm shear}$ from \citet{Maeder1997} and the horizontal turbulence $D_{\rm h}$ from \citet{Zahn1992}. Both prescriptions are standard in Geneva-type models of massive stars up to $120\,M_\odot$.  

The angular momentum distribution is evolved through the Lagrangian form:
\begin{equation}
\rho\,\frac{\partial}{\partial t}\left(r^2\,\Omega\right)_{M_r}
\;=\;
\frac{1}{5\,r^2}\,\frac{\partial}{\partial r}\Big[\rho\,r^4\,\Omega\,U_2(r)\Big]
\;+\;
\frac{1}{r^2}\,\frac{\partial}{\partial r}\Big[\rho\,D_{\rm ang}\,r^4\,\frac{\partial \Omega}{\partial r}\Big],
\label{eq:ang_mom}
\end{equation}
where $\Omega$ is the (shellular) angular velocity, $U_2(r)$ denotes the radial component of the meridional circulation, and $D_{\rm ang}=D_{\rm shear}$ for these non-magnetic models.

Additional physics is captured by the effective diffusion coefficient $D_{\rm eff}$, which accounts for the mixing induced by the interplay of meridional circulation and horizontal turbulence. Our prescription for $D_{\rm shear}$ models the destabilizing effect of differential rotation countered by the stabilizing influence of thermal and compositional gradients, leading to efficient mixing in the radiative zones. Furthermore, the assumption of strong horizontal turbulence smooths out latitudinal variations, allowing a one-dimensional (radial) treatment of the mixing. These rotational transport mechanisms critically affect both the chemical stratification and angular momentum distribution, thereby influencing the evolutionary paths of our supermassive star models. 


Line‐driven winds at $Z=0$ are included using the H and He resonance line prescription of \citet{Kudritzki2002} and \citet{Ekstrom2008} for $\log(L/L_\odot)>6$. The formalism retains the canonical $\dot{M}\!\propto\!(Z/Z_\odot)^{0.5}$ scaling and channels photon momentum through the strongest hydrogen and helium transitions.  
This yields rates far below those of metal-rich counterparts yet provides a physically motivated sink of surface angular momentum. This results in extremely low mass loss rates, where even the most massive 9000 \Ms\ model only loses $\sim$ 2 \Ms\ by the end of the main sequence. Pulsation-driven mass loss is poorly constrained at zero metallicity and is therefore excluded. The models thus evolve at almost constant mass until the surface reaches the $\Omega\Gamma$ limit and rotational–radiative coupling drives outflows.

\subsection{The $\Omega\Gamma$ limit}
Massive stars approaching the Eddington limit experience a reduction of their effective gravity due to the combined effects of centrifugal and radiative forces \citep{Langer1998, Glatzel1998, OGlimit2000}. In our shellular rotation framework the total effective gravity is written as
\begin{equation}
g_{\rm tot}=g_{\rm eff}\,(1-\Gamma),
\label{eq:gtot_geff}
\end{equation}
where $g_{\rm eff} = g_{\rm grav}+g_{\rm rot}$ and $g_{\rm grav}=GM/r^2$ is the gravitational acceleration (with gravitational constant $G$, stellar mass $M$, and local radius $r$), and $g_{\rm rot}$ is the centrifugal acceleration. Here, $\Gamma$ represents the local Eddington factor, which compares the actual radiative flux to its maximum allowed value and may vary with latitude due to gravity darkening.

Setting $g_{\rm tot}=0$ leads to two distinct critical velocities. The first critical velocity is obtained by neglecting radiative forces, so that $g_{\rm eff}=0$, which yields the classical break-up condition,
\begin{equation}
v_{\rm crit,1}=\sqrt{\frac{2\,G\,M}{3\,R_{\rm pb}}}\;,
\label{eq:vcrit1}
\end{equation}
where $R_{\rm pb}$ is the polar radius at break-up. In the presence of significant radiative acceleration, a second solution arises by imposing $g_{\rm eff}\,(1-\Gamma)=0$,
i.e., when $\Gamma=1$. This yields a lower critical velocity that accounts for both rotational and radiative effects. Following \citet{OGlimit2000}, the second critical velocity is 
\begin{equation}
\begin{split}
v_{\rm crit,2}^2 &= \Omega^2\,R_{\rm e}^2(\omega) \\
&=\; \frac{81}{16}\,\frac{\bigl[1-\Gamma_{\max}\bigr]\,GM}{V'(\omega)\,R_{\rm eb}^3(\omega)}\\[1mm]
&=\; \frac{9}{4}\,v_{\rm crit,1}^2\,[1-\Gamma_{\max}]\,
\frac{R_{\rm eb}^2(\omega)}{R_{\rm e}^2(\omega)}\;,
\end{split}
\label{eq:vcrit2}
\end{equation}
where $\Omega$ is the angular velocity; $R_{\rm e}(\omega)$ is the equatorial radius for a dimensionless rotation parameter $\omega$; $R_{\rm eb}(\omega)$ is the equatorial radius at break-up; $\Gamma_{\max}$ is the maximum Eddington factor over the stellar surface; and $V'(\omega)$ is a dimensionless function of $\omega$ (see \citealt{OGlimit2000} for further details). In the $\Omega\Gamma$ limit, defined by the condition $\Gamma=1$, even moderate rotation drives the star to break up, leading to dramatically enhanced mass loss. This second solution, $v_{\rm crit,2}$, becomes physically relevant for stars with $\Gamma_{\max}\gtrsim0.639$, a regime typical of our supermassive star models.

In our models, when the surface equatorial velocity, $v_{eq}$, approaches this second critical velocity, $v_{crit,2}$, the star enters a phase of intense mechanical mass loss. This process is not treated with a continuous wind prescription but is implemented as a numerical enforcement of the physical limit. The code continuously monitors the ratio $v_{eq}/v_{crit,2}$. If this ratio reaches unity, a mass-loss routine is activated that removes the necessary amount of mass ($\Delta M$) and angular momentum from the surface layer in that timestep to ensure the star remains just below the critical limit in the subsequent step. The mass loss rate is thus dynamically calculated by the code to enforce the condition:
\begin{equation}
    \dot{M} =
    \begin{cases}
      \dot{M}_{\mathrm{needed}} & \text{if } v_{eq} \geq v_{crit,2} \\
      0 & \text{if } v_{eq} < v_{crit,2}
    \end{cases}
\end{equation}
where $\dot{M}_{\mathrm{needed}}$ is the rate required to maintain the star at the stability boundary. This self-regulating mechanism effectively caps the surface rotation of the star.

\section{Results}\label{Sec:Results}
We show the evolution of the three SMSs at $v/v_{crit} =$ 0.01, 0.05 and 0.1 in the HR diagrams in panels (a), (b), and (c) of Figure~\ref{fig:HRDall} respectively. Total lifetimes (column 3), times spent as blue supergiants (column 4) and red supergiants (column 5) are shown in Table~\ref{tab:popiii_evolution} along with convective core mass fractions (columns 6 and 10), equatorial velocities (columns 7 and 11), total angular momenta (columns 8 and 12) and Eddington factors (columns 9 and 13).

\subsection{Evolution in the HR diagram}\label{Sec:HR}
All three sets of models begin their lifetimes as hot and compact blue supergiants with stellar stellar radii of 16 - 55 \Rs\ and an effective temperature around 88,000 K. Following a short contraction phase that lasts a few thousand years, all models reach a central temperature of log ($T_{\mathrm{eff}}$) = 8.15, marking the start of core hydrogen burning. All models possess large convective cores that make up about 95\% of the total star in mass. As hydrogen continues to fuse in the convective core, the total energy generated is initially dominated by the pp chain. However, the central temperature is high enough for the 3-$\alpha$ reaction to produce $^{12}$C and once its central abundance exceeds a mass fraction of 10$^{-10}$, the CNO cycle is initiated. The energy generation from CNO burning contributes significantly to the total energy production of the star, and this energy is transported from the center to the surface. A fraction of this energy goes into in expanding the envelope, thereby increasing the stellar radius. As the models continue to deplete their central hydrogen reservoir, their stellar radius increases as they steadily begin their transition into red supergiant protostars with stellar radii varying from 120 - 10,000 \Rs. 

The expansion of the stars naturally affects their equatorial velocities throughout their evolution. For instance, the equatorial velocities range between 26 - 44 km s$^{-1}$ at the start of core hydrogen burning, making these particular models slow rotators in spite of their compact size. This is simply due to the choice of initial rotation, which is based on the $\Omega\Gamma$ limit and will be explored in a later section. These equatorial velocities later fall to nearly zero, especially in case of more massive models (5000 - 10,000 \Ms) as the central hydrogen is depleted. An interesting feature to note here is the position of the tracks in the HR diagram at the end of core hydrogen burning. We find that as SMS mass increases, core hydrogen burning ends in a more red-ward position in the HR diagram. This result is in agreement with standard rotating 9 - 120 \Ms\ star models. The main-sequence lifetime of these stars ranges from 1.0 Myr (10,000 \Ms) - 1.6 Myr (1000 \Ms), as shown in column 2 of Table~\ref{tab:popiii_evolution}. Although the mass difference between the two models is one order of magnitude (1000\%), the difference in age is only 60\%. This is a due to a shift in the luminosity-mass relationship, where the scaling shifts to L $\propto$ M. In addition, all models spend the majority of their lifetimes (90 - 100\%) on the blue side of the HR diagram (log ($T_{\mathrm{eff}}$) $>$ 4), as seen in columns 4 and 5 of Table~\ref{tab:popiii_evolution}. 

\begin{figure*}[!htbp]
  \centering
  \includegraphics[width=\textwidth]{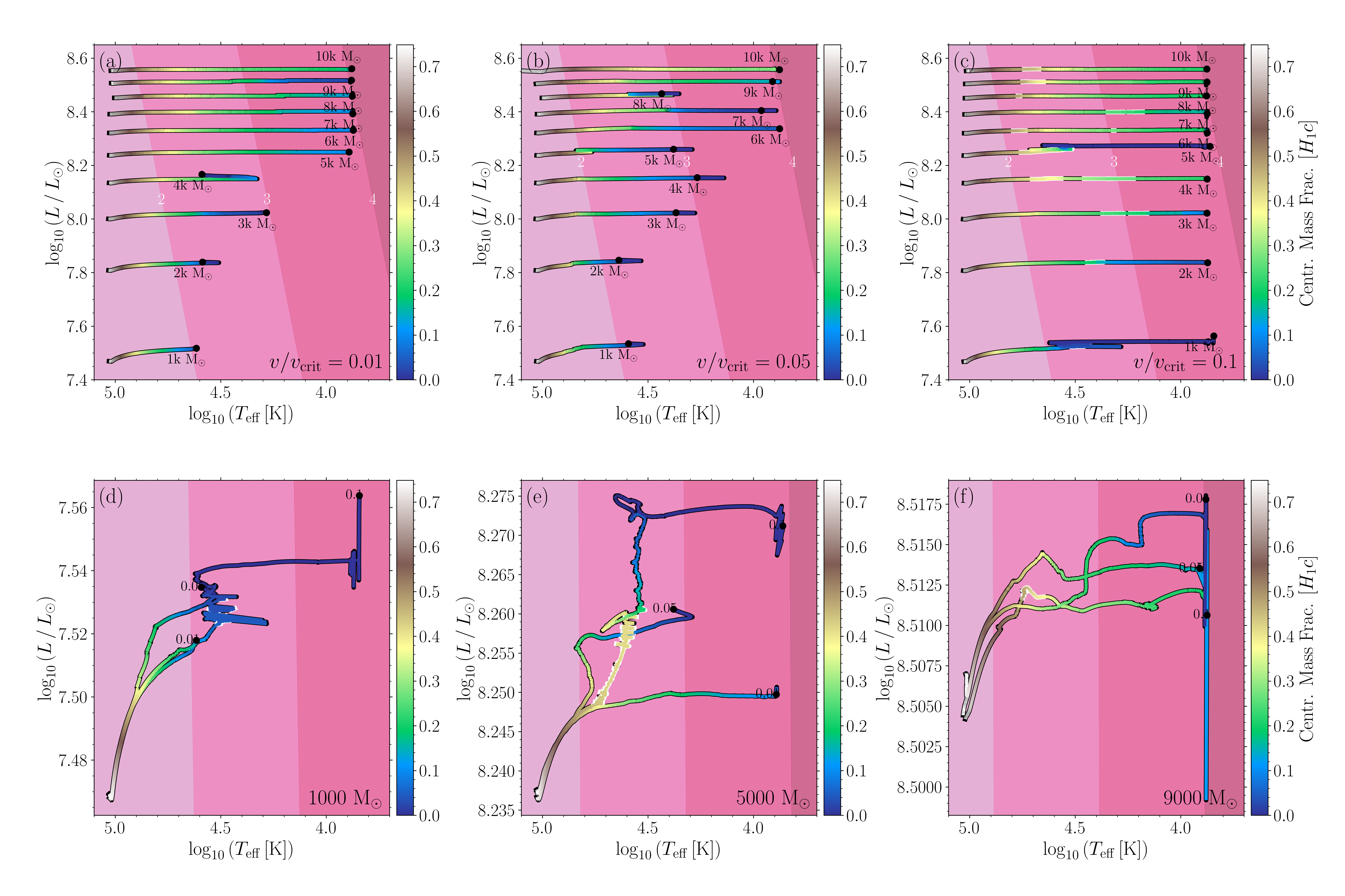}
  \caption{SMS evolution on the HR diagram. The colourbar indicates  central H mass fractions, $X_{\rm c}$, and the shaded red regions show the isoradii lines, with the lightest shade corresponding to log (R/\Rs) = 1 and darkest shade depicting log (R/\Rs) = 4. The white outline around stellar tracks correspond to the evolutionary stage where the $\Omega\Gamma$‑limit is encountered. {\it Top row:} 1000 - 10000 \Ms\ stars at initial rotation rates $v/v_{\mathrm{crit}}=$  0.01 (a), 0.05 (b) and 0.10 (c). {\it Bottom row:} Comparison of rotation rates at a given mass, 1000 \Ms\ (d), 5000 \Ms\ (e), and 9000 \Ms\ (f).}
  \label{fig:HRDall}
\end{figure*}

As shown in panel (b) of Figure~\ref{fig:HRDall}, the 0.05 $v/v_{\mathrm{crit}}$ SMS exhibits evolutionary trends in nuclear reaction rates, effective temperatures, and convective core properties that are quite similar to those of the 0.01 $v/v_{\mathrm{crit}}$ model. The slight but important differences arise in the total main sequence lifetimes, as they range from 1 Myr (10,000 \Ms) - 1.74 Myr (1000 \Ms), as shown in column 2 of Table~\ref{tab:popiii_evolution}. This increase in lifetime on the lower mass end, for instance at 1000 \Ms, is around 7\% larger than the same star at the lowest rotation rate. This effect is primarily due to stronger elemental mixing  (in this case hydrogen) during the main sequence made possible by the increased rotation rate. Hydrogen from the radiative envelope is dredged down to the convective core by rotational mixing facilitated by meridional currents, which partially resupplies the core and extends its size. This can be seen in columns 6 and 10 of Table~\ref{tab:popiii_evolution}, where the convective core mass fraction, $M_{\mathrm cc}$, is larger for the 0.05 $v/v_{\mathrm{crit}}$ models than for the corresponding 0.01 $v/v_{\mathrm{crit}}$ models, especially by the end of core hydrogen burning. Because of the higher rotation rate, the equatorial velocities (columns 7 and 11 of Table~\ref{tab:popiii_evolution}) are naturally higher for 0.05 $v/v_{\mathrm{crit}}$ models at the onset of core hydrogen burning, but this trend is not seen at the end of core hydrogen burning in the 5000 - 10,000 \Ms\ models. These stars also approach the Hayashi limit and their radii, and therefore their equatorial velocities, are identical to those of the 0.01 $v/v_{\mathrm{crit}}$ models. The total angular momentum   $L_{\mathrm{ tot}}$ reservoirs (columns 8 and 12 of Table~\ref{tab:popiii_evolution}) of these models are also higher throughout their evolution. A point of similarity between the two sets of models here is the Eddington factor ($\Gamma_{\mathrm{edd}}$), shown in columns 9 and 13 of Table~\ref{tab:popiii_evolution}. All models, independent of rotation rates, exhibit very high $\Gamma_{\mathrm{edd}}$, which makes them susceptible to the $\Omega\Gamma$ limit (see section~\ref{Sec:OG}). 

The general evolutionary trends of the 0.1 $v/v_{\mathrm{crit}}$ model shown in panel (c) of Figure~\ref{fig:HRDall} are again similar to those at 0.01 and 0.05 $v/v_{\mathrm{crit}}$, with a few additional insights. The extent of chemical mixing in these models is highest among the three rotation rates and produces the largest $M_{\mathrm{cc}}$. This in turn extends their lifetimes by another few percent when compared to the 0.05 $v/v_{\mathrm{crit}}$ models. Stronger mixing also boosts energy production in the stars. As this energy propagates through the stellar interior it expands the envelope to even greater radii than those of the other two sets of models, ensuring that stars at these highest rotation rates spend the longest times in the red (column 5 of Table~\ref{tab:popiii_evolution}) and subsequently end the main sequence on the red side of HR diagram. Although these models have the highest initial equatorial velocities and total angular momentum reservoirs, these velocities drop to zero once they approach the Hayashi line and become red supergiants. 

We compare the evolution of 1000 \Ms, 5000 \Ms, and 9000 \Ms\ stars at all three rotation rates in panels (d), (e), and (f) of Figure~\ref{fig:HRDall}. 
The effect of rotational mixing is evident in the 1000 \Ms\ HR tracks when they begin to diverge after beginning the main sequence at the same place in the HR diagram. The 0.01 $v/v_{\mathrm{crit}}$ star exits the main sequence with log ($L/L_{\odot}) =$ 7.51 and log ($T_{\mathrm{eff}} =$ 4.6. As chemical mixing is enhanced by higher rotation rates, the 0.05 $v/v_{\mathrm{crit}}$ star becomes slightly more luminous, log ($L/L_\odot) =$  7.53 and cooler, log ($T_{\mathrm{eff}}) =$ 4.5. This effect is enhanced in the 0.1 $v/v_{\mathrm{crit}}$ model, where log ($L/L_\odot) =$ 7.56 and log ($T_{\mathrm{eff}})$ = 3.8. At log ($L/L_\odot) =$ 7.525, the 0.1 $v/v_{\mathrm{crit}}$ model fluctuates between the red and blue sides of the HR diagram because $\Gamma_{\mathrm{edd}}$ exceeds 0.9, causing the star to encounter the $\Omega\Gamma$ limit. The consequences of this limit will be discussed later. 

The effects of different rotation rate are also evident in the 5000 \Ms\ models in panel (e) of Figure~\ref{fig:HRDall}, as the luminosities vary by $\sim$ log ($L/L_\odot$) = 0.075. The effective temperature of the 0.05 $v/v_{\mathrm{crit}}$ SMS is much higher than for the other two models, as it ends its core hydrogen burning stage at log ($T_{\mathrm{eff}}$) = 4.4. Generally, this effect can be attributed to stronger mixing in the convective core induced by rotation and this fact is true when comparing the 0.01 $v/v_{\mathrm{crit}}$ and 0.1 $v/v_{\mathrm{crit}}$ models. The 0.1 $v/v_{\mathrm{crit}}$ star is indeed  redder because of stronger rotational mixing, which allows more efficient transport of hydrogen into the core and builds up the helium core at a higher rate. This raises the luminosity even higher and the model approaches the Eddington luminosity sooner. But 0.05 $v/v_{\mathrm{crit}}$ being less red than 0.01 $v/v_{\mathrm{crit}}$ could be attributed to the core properties and temperature sensitivity of the CNO cycle. In the case of 0.05 $v/v_{\mathrm{crit}}$ the initial production of $^{12}$C by 3$\alpha$ is delayed, which in turn affects energy production by the CNO cycle. The 0.1 $v/v_{\mathrm{crit}}$ star also encounters the $\Omega\Gamma$ limit at log ($L/L_\odot$) = 8.256.

Finally, for the 9000 \Ms\ models shown in panel (d) of Figure~\ref{fig:HRDall}), the trends are also quite similar to those of the 1000 \Ms\ models but the difference in luminosities in the fastest versus slowest rotating models is even smaller, log ($L/L_\odot$) = 0.007. At such high masses, all models reach the red side of the HR diagram before the end of core hydrogen burning, with near-zero equatorial velocities. The value of $\Gamma_{\mathrm{edd}}$ is also quite similar for the three rotation rates, about 0.965. The 0.05 $v/v_{\mathrm{crit}}$ model is on the verge of the $\Omega\Gamma$ limit and 0.1 $v/v_{\mathrm{crit}}$ model encounters it at log ($L/L_\odot$) $=$ 8.514.

\rowcolors{2}{gray!15}{gray!05}
\begin{table*}[htbp]
  \centering
  \scriptsize
  \renewcommand{\arraystretch}{0.9}
  \resizebox{\textwidth}{!}{%
    \begin{tabular}{c c c c c c c c c c c c c}
      \toprule
      \multicolumn{5}{c}{} 
        & \multicolumn{4}{c}{Start of H burning}
        & \multicolumn{4}{c}{End of H burning} \\
      Mass & $v/v_{\mathrm{crit}}$ & lifetime & $t_{\rm blue}$ & $t_{\rm red}$
        & $M_{\rm cc}$ & $v_{\rm eq}$ & $L_{\rm tot}$ & $\Gamma_{\rm Edd}$
        & $M_{\rm cc}$ & $v_{\rm eq}$ & $L_{\rm tot}$ & $\Gamma_{\rm Edd}$ \\
      (\Ms) &  & (Myr) & (Myr) & (Myr)
        &  & (km s$^{-1}$) & (10$^{53}$ g cm$^2$ s$^{-1}$) & 
        &  & (km s$^{-1}$) & (10$^{53}$ g cm$^2$ s$^{-1}$) &  \\
      \midrule
      1000  & 0.01 & 1.658 & 1.637 & 0.000 &  0.942 &  26.600 &   5.108 & 0.793 &  0.510 &   7.360 &   4.930 & 0.883 \\
      2000  & 0.01 & 1.442 & 1.426 & 0.000 &  0.952 &  31.400 &  16.040 & 0.865 &  0.510 &   2.360 &  16.044 & 0.923 \\
      3000  & 0.01 & 1.350 & 1.335 & 0.000 &  0.955 &  34.500 &  31.402 & 0.898 &  0.506 &   9.950 &  30.754 & 0.942 \\
      4000  & 0.01 & 1.286 & 1.273 & 0.000 &  0.955 &  36.700 &  49.478 & 0.917 &  0.520 &  14.800 &  46.923 & 0.950 \\
      5000  & 0.01 & 1.175 & 1.106 & 0.056 &  0.957 &  38.700 &  71.264 & 0.930 &  0.564 &   3.000 &  66.299 & 0.962 \\
      6000  & 0.01 & 1.168 & 1.086 & 0.070 &  0.958 &  40.400 &  96.316 & 0.940 &  0.608 &   3.000 &  80.615 & 0.954 \\
      7000  & 0.01 & 1.164 & 1.043 & 0.109 &  0.959 &  41.800 & 122.912 & 0.947 &  0.480 &   3.000 & 108.858 & 0.946 \\
      8000  & 0.01 & 1.117 & 1.001 & 0.104 &  0.959 &  41.000 & 132.190 & 0.953 &  0.568 &   0.144 & 119.633 & 0.962 \\
      9000  & 0.01 & 1.143 & 1.099 & 0.037 &  0.958 &  44.100 & 177.906 & 0.959 &  0.511 &   0.146 & 175.794 & 0.971 \\
      10000 & 0.01 & 1.028 & 0.901 & 0.121 &  0.957 &  45.100 & 209.965 & 0.963 &  0.640 &   0.305 & 162.181 & 0.957 \\
      \midrule
     1000   & 0.05 & 1.741 & 1.720 & 0.000 &  0.941 & 134.000 &  25.691 & 0.793 &  0.522 &  25.600 &  25.671 & 0.896 \\
     2000   & 0.05 & 1.463 & 1.447 & 0.000 &  0.950 & 158.000 &  80.666 & 0.865 &  0.524 &  24.100 &  80.759 & 0.926 \\
     3000   & 0.05 & 1.348 & 1.334 & 0.000 &  0.954 & 174.000 & 158.032 & 0.897 &  0.527 &   0.311 & 158.876 & 0.942 \\
     4000   & 0.05 & 1.286 & 1.273 & 0.000 &  0.954 & 184.000 & 248.373 & 0.916 &  0.543 &   0.000 & 249.705 & 0.953 \\
     5000   & 0.05 & 1.260 & 1.247 & 0.000 &  0.955 & 194.000 & 357.462 & 0.929 &  0.530 &   2.100 & 359.237 & 0.961 \\
     6000   & 0.05 & 1.227 & 1.158 & 0.058 &  0.957 & 203.000 & 485.122 & 0.939 &  0.571 &   0.991 & 446.133 & 0.970 \\
     7000   & 0.05 & 1.204 & 1.177 & 0.015 &  0.956 & 210.000 & 617.091 & 0.947 &  0.572 &   1.870 & 619.190 & 0.970 \\
     8000   & 0.05 & 1.178 & 1.166 & 0.000 &  0.958 & 206.000 & 658.331 & 0.953 &  0.575 &   0.001 & 662.250 & 0.970 \\
     9000   & 0.05 & 1.158 & 0.977 & 0.013 &  0.959 & 230.000 & 970.288 & 0.958 &  0.646 &   0.000 & 975.517 & 0.971 \\
    10000   & 0.05 & 1.070 & 0.901 & 0.103 &  0.967 & 179.000 & 940.312 & 0.960 &  0.709 &   1.720 & 792.292 & 0.963 \\
      \midrule
     1000   & 0.10 & 1.848 & 1.768 & 0.059 &  0.939 & 268.000 &  51.182 & 0.791 &  0.523 & 28.005 &  46.985 & 0.906 \\
     2000   & 0.10 & 1.505 & 1.440 & 0.058 &  0.947 & 315.000 & 160.639 & 0.863 &  0.556 &   0.748 & 127.237 & 0.937 \\
     3000   & 0.10 & 1.413 & 1.266 & 0.133 &  0.950 & 347.000 & 314.233 & 0.895 &  0.613 &   0.768 & 261.266 & 0.949 \\
     4000   & 0.10 & 1.322 & 1.202 & 0.107 &  0.952 & 368.000 & 492.497 & 0.915 &  0.671 &   1.110 & 390.916 & 0.958 \\
     5000   & 0.10 & 1.300 & 1.285 & 0.002 &  0.954 & 383.000 & 705.955 & 0.928 &  0.550 & 2.802 & 591.312 & 0.965 \\
     6000   & 0.10 & 1.315 & 1.163 & 0.139 &  0.954 & 406.000 & 961.128 & 0.937 &  0.051 &   0.014 & 779.156 & 0.947 \\
     7000   & 0.10 & 1.236 & 1.094 & 0.129 &  0.954 & 420.000 &1225.720 & 0.945 &  0.243 &   1.550 &1035.676 & 0.951 \\
     8000   & 0.10 & 1.270 & 1.123 & 0.135 &  0.955 & 434.000 &1511.272 & 0.951 &  0.619 &   1.550 &1289.506 & 0.971 \\
     9000   & 0.10 & 1.227 & 1.097 & 0.122 &  0.956 & 448.000 &1817.557 & 0.956 &  0.657 &   3.370 &1530.294 & 0.965 \\
    10000   & 0.10 & 1.148 & 0.975 & 0.178 &  0.955 & 449.000 &2084.810 & 0.961 &  0.710 &   5.460 &1765.897 & 0.965 \\
      \bottomrule
    \end{tabular}%
  }
  \caption{Evolutionary parameters for Pop\,III SMS models:
    lifetime and cumulative time in red ($\log T_{\rm eff}\le4.00$) and blue ($\log T_{\rm eff}>4.00$) phases;
    core mass fraction $M_{\rm cc}$, equatorial velocity $v_{\rm eq}$, total angular momentum $L_{\rm tot}$,
    and Eddington factor $\Gamma_{\rm Edd}$ at the start and end of H burning.}
  \label{tab:popiii_evolution}
\end{table*}

\subsection{Rotation and transport of chemical species}\label{Sec:Chem}
\begin{figure*}[!htbp]
    \centering\includegraphics[width=\textwidth]{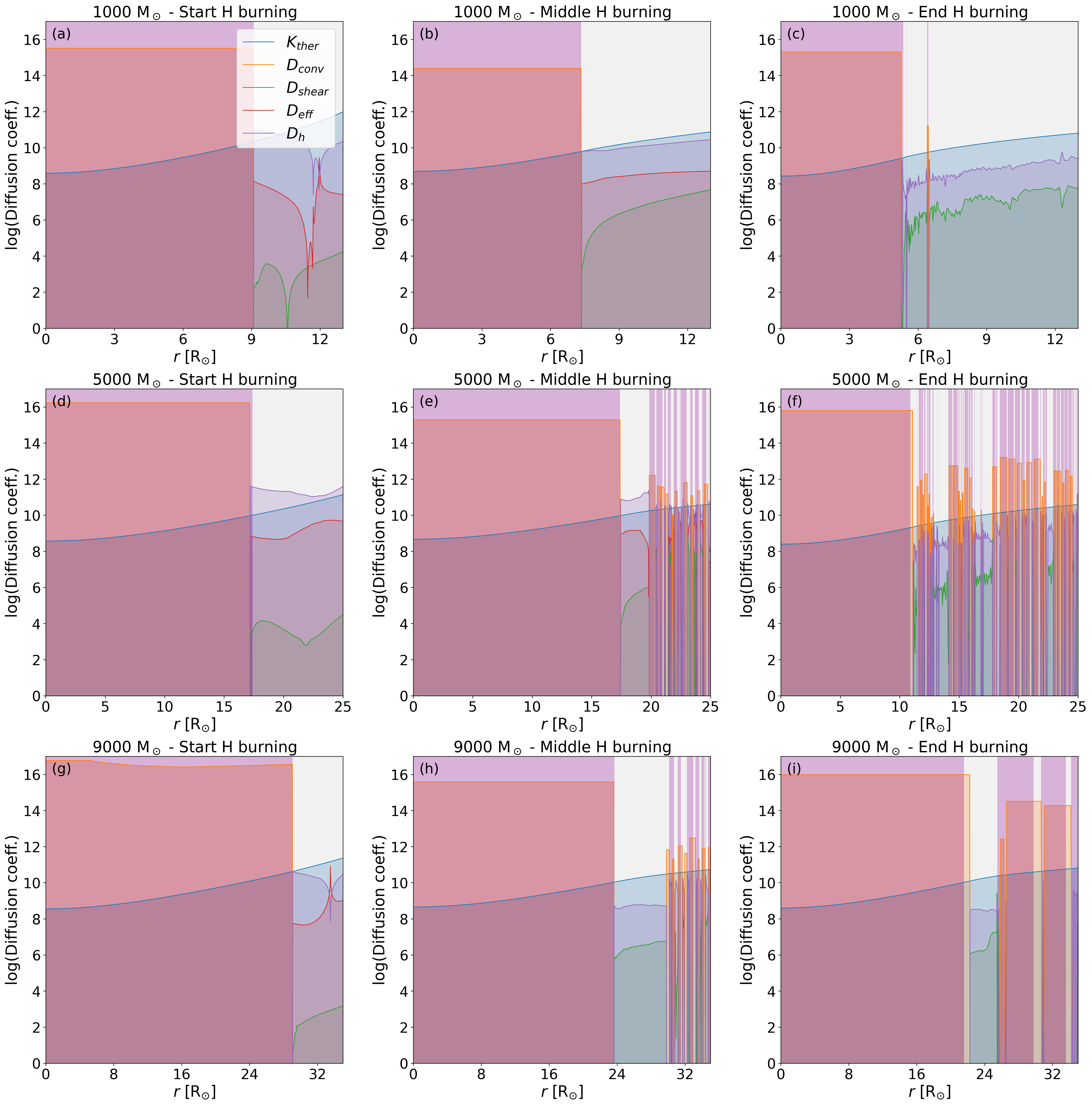}
    \caption{Evolution of diffusion coefficients in the 0.1 $v/v_{\mathrm{crit}}$ models. The lavender and off-white regions show convective and radiative zones, respectively. Multiple diffusion coefficients overlap a given zone, but the dominant coefficient is the one with the highest magnitude along the y-axis. $D_{\rm conv}$ is the convective diffusion coefficient, $K_{ther}$ is the thermal diffusivity coefficient, $D_{\rm shear}$ is the shear (vertical) mixing coefficient, $D_{\rm h}$ is the horizontal mixing coefficient, and $D_{\rm eff}$ is the effective mixing coefficient. {\it Top row:} the 1000 \Ms\ model at the onset, middle, and end of core hydrogen burning. {\it Center row:} the 5000 \Ms\ model at the beginning, middle, and end of core hydrogen burning. {\it Bottom row:} the 9000 \Ms\ model at the start, middle, and end of core hydrogen burning.}
    \label{fig:DiffCoeff}
\end{figure*}

We now consider the impact of diffusion coefficients on  stellar structure at different stages of evolution. The impact of rotation will be most evident in the 0.1 $v/v_{\mathrm{crit}}$ models, so we show the evolution of these coefficients in the 1000, 5000 and 9000 \Ms\ stars at this rate in Figure~\ref{fig:DiffCoeff}. All panels (a) - (i) depict the inner zones of these models with focus on the boundary of convective core and radiative envelope. 

At the start of core hydrogen burning in panels (a), (d), (g), we find that $D_{\rm conv}$ is the dominant mixing coefficient in the convective core and the mixing of chemical species in this zone is assumed to be nearly instantaneous. The radii of the convective cores for all three stars shows an intuitive trend, where the 9000 \Ms\ model has the largest (29 \Rs), followed by 5000 \Ms\ (17.5 \Rs) and 1000 \Ms\ (9 \Rs). The thermal diffusivity coefficient, $K_{\mathrm{ther}}$, is responsible for damping temperature perturbations on a timescale $l^{2}/K$. In the context of shear instabilities, a large $K_{\mathrm{ther}}$ lowers the effective Richardson number by smoothing out buoyancy forces. Hence the “secular shear” develops when thermal diffusion is fast compared to the turnover of shear eddies. In context of these models, higher $K_{\mathrm{ther}}$ (more efficient radiative diffusion) enhances mixing by weakening stabilizing thermal gradients, and this is evident in radiative zones for all three models throughout the main sequence. 

At the boundary of the convective core with a strong mean molecular weight gradient, $\nabla_{\mu}$, we find that the horizontal mixing coefficient $D_{\rm h}$ is the strongest in magnitude. As seen in Equation \ref{eq:Dh}, $D_{\rm h}$ is inversely proportional to $D_{\rm eff}$, the effective mixing coefficient, but in spite of this dependence it is $D_{\rm eff}$ that determines  chemical transport at the boundary of convective core. This is because $D_{\rm eff}$ is directly proportional to the radial amplitude of the meridional circulation velocity, and circulation is driven by the thermal imbalance (baroclinicity) created by centrifugal distortion and the resulting horizontal temperature gradients. These temperature gradients are strongest at the convective - radiative boundary, which leads to a much larger meridional circulation velocity than $D_{h}$ that results in chemical transport dominated by $D_{\rm eff}$. Moving outward in the radiative regions, the chemical transport is governed by the vertical mixing coefficient $D_{\rm shear}$ all the way to the surface of all models. All three models at this evolutionary stage have identical stellar structures and all the mixing coefficients behave in the same manner at all masses. 

The stellar structures of the three models begin to show differences at the middle of core hydrogen burning. For the 5000 and 9000 \Ms\ stars (panels (e), and (h) in Figure~\ref{fig:DiffCoeff}), intermediate convective zones are formed starting at radii of 20 and 30 \Rs, respectively. These intermediate convective zones are formed as a result of the radiative temperature gradient, $\nabla_{\rm rad}$, exceeding the adiabatic temperature gradient,  $\nabla_{\rm ad}$, in the local region. The increase in $\nabla_{\rm rad}$ is likely due to the change in local opacity, energy generation rate, and mean molecular weight profile. As in the convective core,  chemical transport in these intermediate convective zones is dominated by $D_{\rm conv}$. Every other mixing coefficient trend is identical to that at the onset of core hydrogen burning. 

At the end of core hydrogen burning (panels (c), (f), and (i) in Figure~\ref{fig:DiffCoeff}), the convective core in all models has receded and helium is being accumulated in the core. The 5000 and 9000 \Ms\ stars are now dominated by intermediate convective zones all the way to the surface, making these models 70-90\% convective. At this stage in evolution, transport of chemical species is largely done by $D_{\rm conv}$. Conclusively, these results show that $D_{\rm eff}$-driven meridional circulation dominates chemical transport at the convective–radiative boundary. Subsequent radiative regions are mixed by $D_{\rm shear}$ which yields a more uniform envelope composition that can influence burning lifetimes and core growth. Since most of the stellar structure is convective, the transport of chemical species is dominated by the physics of convection.

\subsection{Rotation and transport of angular momentum}\label{Sec:Ang}
Rotation also impacts the transport of angular momentum in these stars, but not directly via the diffusion coefficients. In these models (see Eq.~\ref{eq:ang_mom}), the dominant mechanism for angular‐momentum transport is the advective flux associated with meridional circulation, quantified by the radial velocity component $U_2(r)$. Shear instabilities, through the diffusion coefficient $D_{\rm shear}$, provide a complementary, diffusive redistribution of angular momentum. Horizontal turbulence, $D_h$, acts to enforce nearly uniform angular velocity on isobars but contributes minimally to the net transport. Together, these processes establish a balance between advection and diffusion, enabling efficient angular‐momentum redistribution in  SMSs.  

\subsection{Supermassive stars and the $\Omega\Gamma$ limit}\label{Sec:OG}

\begin{figure}[!h]
  \centering
  \includegraphics[width=9cm]{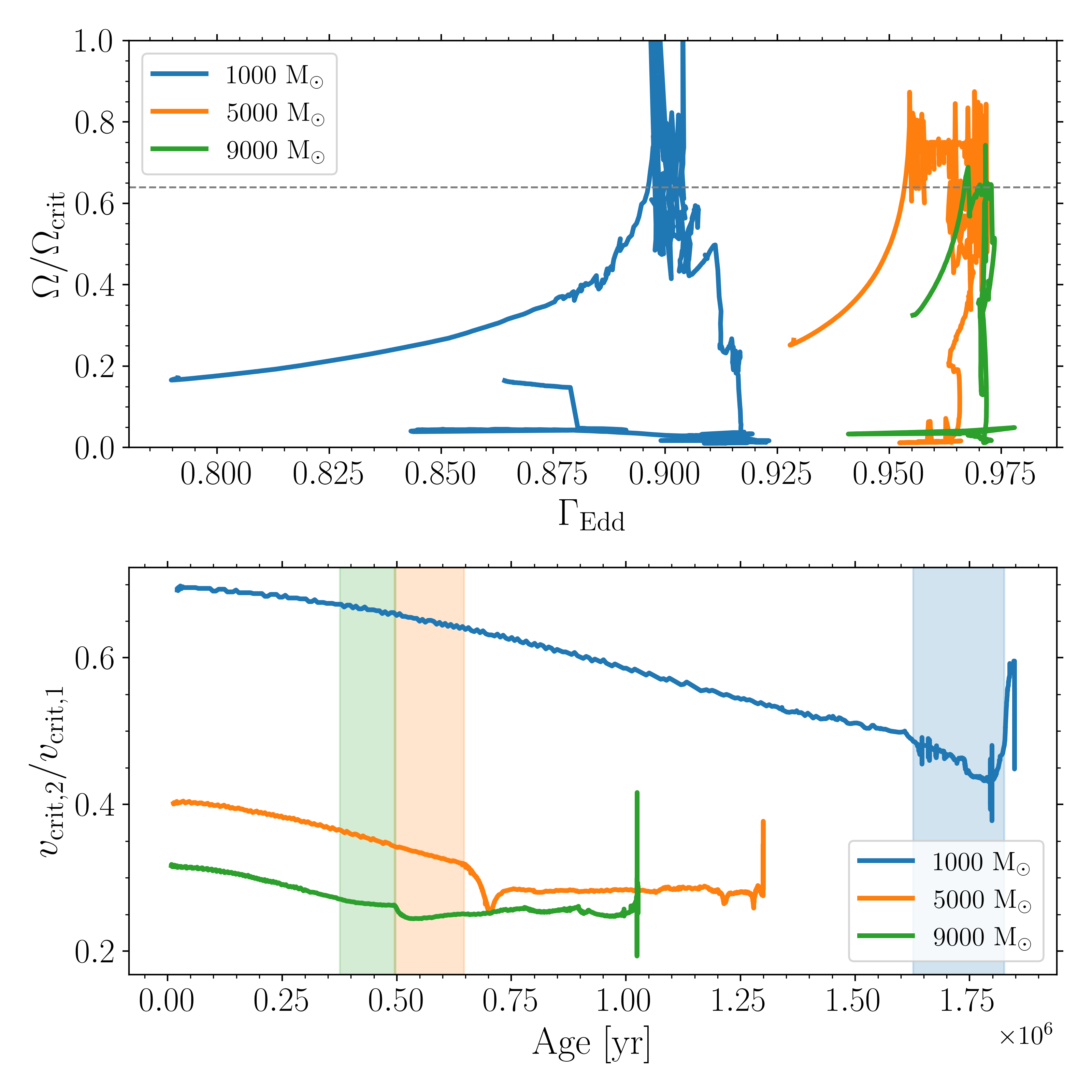}
  \caption{Effects of the $\Omega\Gamma$ limit on the 1000, 5000, and 9000 \Ms\ models at 0.1 $v/v_{\mathrm{crit}}$. {\it Top:} Evolution of the surface spin parameter $\Omega/\Omega_{\rm crit}$ versus Eddington factor $\Gamma$. The horizontal dashed line marks $\Gamma=0.639$, above which the $\Omega\Gamma$‑limit (Eq.~\ref{eq:vcrit2}) becomes relevant.  
  {\it Bottom:} Ratio of radiatively reduced to classical critical velocity, $v_{\rm crit,2}/v_{\rm crit,1}$, versus stellar age.  The decline from the initial values of $\sim$ 0.69, 0.40, and 0.32 to minima of $\sim$ 0.38, 0.25 and 0.19 quantifies the suppression of break‑up speed by radiative acceleration. The shaded zones represent the age where $\Omega\Gamma$‑limit is active.}
  \label{fig:OOc_vs_Gamma}
\end{figure}

In order to quantify the effect of the $\Omega\Gamma$‑limit on our SMS models, we focus first on the set with the highest initial rotation ($v/v_{\rm crit}=0.1$), which provides an upper bound on rotationally induced effects.  Figure~\ref{fig:OOc_vs_Gamma} (top panel) shows the evolution of the surface spin parameter $\Omega/\Omega_{\rm crit}$ as a function of the local Eddington factor $\Gamma$.  All three masses start the main sequence at substantial fractions of break‑up ($\Omega/\Omega_{\rm crit}\simeq0.17,0.26,0.33$ at $\Gamma\simeq0.79,0.93,0.96$ respectively) and rapidly spin up. The value for $\Omega/\Omega_{\rm crit}$ reaches $\simeq0.7$ when the ages of the 1000, 5000, and 9000 \Ms\ stars are $2.8\times10^4\,$yr, $7.0\times10^5\,$yr, and $8.2\times10^5\,$yr, respectively, beyond which the centrifugal acceleration $g_{\rm rot}$ rivals the reduced gravity $g_{\rm grav}(1-\Gamma)$ and mechanical winds must ensue \citep[e.g.][]{Langer1998}. After peaking (up to $\Omega/\Omega_{\rm crit}\simeq0.87$ and $0.74$ for the two highest masses), all models spin down as mass and angular momentum are shed.

Figure~\ref{fig:OOc_vs_Gamma} (bottom panel) plots the ratio $v_{\rm crit,2}/v_{\rm crit,1}$ as a function of age.  This ratio follows 
\[
  \frac{v_{\rm crit,2}}{v_{\rm crit,1}}
  =\frac{3}{2}\,\sqrt{1-\Gamma(t)}\,\frac{R_{\rm eb}}{R_e}\,,
\]
and falls from $\sim 0.69, 0.40, 0.32$ at zero‐age to minima of $\sim 0.38, 0.25, 0.19$ near each model’s maximum $\Gamma$ (Eqs.~\ref{eq:vcrit1}–\ref{eq:vcrit2}).  The more massive stars, with $\Gamma_{\max}\gtrsim0.96$, suffer the strongest reduction of their effective break‑up speed, demonstrating that even moderate rotation forces them well into the $\Omega\Gamma$‑limit regime and amplifies mass‐loss rates by orders of magnitude \citep{OGlimit2000}. To determine a lower boundary in rotation for the $\Omega\Gamma$ limit, we repeated this analysis for the $v/v_{\rm crit}=0.05$ runs.  We find that models in the 8000–10000 \Ms\ range cross $\Omega/\Omega_{\rm crit}\simeq0.7$ during the main sequence as well, indicating that mechanical mass loss will set in even at this rotation rate.  

In summary, the $\Omega\Gamma$‑limit imposes a stringent cap on SMS rotation: any approach to $\Omega/\Omega_{\rm crit}\gtrsim0.7$ triggers strong, mechanical mass loss that prevents sustained high‐spin states.  This regulatory mechanism, initially proposed for massive stars over a range of metallicities applies equally to supermassive Pop III stars, constraining their angular momenta and influencing their ultimate fates \citep{OGlimit2000}. However, the extent of mass loss needed to surpass this limit is quite low in numerical simulations. For instance, the 9000 \Ms\ star at 0.1 $v/v_{\mathrm{crit}}$ only loses 147 \Ms\ over its life. This mass, although significant, is only 1.6\% of its total mass. To better quantify mass loss, full 3D hydrodynamic simulations for mass loss outflows from SMSs are required. 

\subsection{Estimates on the core rotation rates}\label{Sec:Core}

\begin{figure*}[!htbp]
  \centering
  \includegraphics[%
    width=\textwidth,%
    height=0.75\textheight,%
    keepaspectratio%
  ]{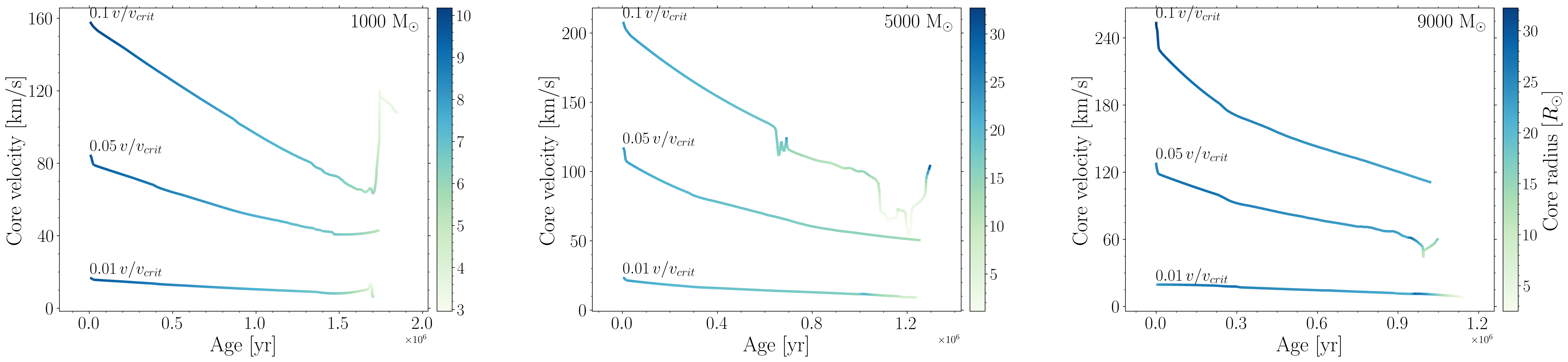}
  \caption{Convective‑core rotation velocity versus age for models at three rotation rates. The second y axis shows the radius of the core in solar radii. {\it Left:} 1000 \Ms\ models, {\it Centre:} 5000 \Ms\ models, {\it Right:} 9000 \Ms\ models.}
  \label{fig:corevel}
\end{figure*}

Figure~\ref{fig:corevel} complements our HR‑diagram discussion by showing the mass–weighted rotation velocity of the convective core, $v_{\mathrm c}$, from ZAMS to the end of core H burning for our three Pop III masses. Each panel contains the tracks that start with $v/v_{\mathrm{crit}}=0.01,\;0.05,$ and $0.10$; the colour scale gives the instantaneous core radius $R_{\mathrm c}$. For an initial rotation of $v/v_{\mathrm{crit}}=0.1$, the 1000, 5000, and 9000 \Ms\ models decelerate almost linearly from initial core rotational velocity $\simeq$ 160, 210, and 250 km s$^{-1}$, respectively, to 40--70, 50--80, and 60--90 km s$^{-1}$ by the end of core H burning. Metal-free winds are negligible, so the loss of core angular momentum must be mediated by internal transport: meridional advection extracts $J$ from the fully convective region and feeds it into the differentially rotating envelope, overcoming the modest contraction of $R_{\mathrm c}$. The decline rate steepens with mass because the moment of inertia grows faster when larger fractions of the star participate in convection.

At fixed mass the vertical spacing of the three tracks reflects the imposed surface spin. After an adjustment phase lasting $\sim(0.1$--$0.15) \, t_{\mathrm{MS}}$ the slopes converge, signalling that the core has attained near solid‑body rotation and evolves thereafter under the same angular‑momentum sink, with a similar rate of change of core velocity with time ($dv_c/dt$). A global fit to the combined data gives
$v_{\mathrm c}\propto R_{\mathrm c}^{0.3\pm0.1}$, consistent across all masses and rotation rates and confirming that the decay of $J_{\mathrm c}$ dominates over structural changes.

The $5\times10^{3}$ \Ms, 0.10 $v/v_{crit}$ model deviates briefly from the other stars from $t\simeq5.5\times10^{5}$ and $9\times10^{5}\,\mathrm{yr}$: a short-lived convective shell at $r\simeq25$ \Rs\ drains angular momentum from the core, flattens the $v_{\mathrm c}(t)$ slope, and disappears once the local $\mu$-gradient is re-established. Model diagnostics show that many intermediate convective shells form at the core–envelope interface; the shell taps angular momentum from the core, flattens the $v_{\mathrm c}(t)$ curve, and disappears once the local $\mu$ gradient is restored, after which the track rejoins the general trend.

Hydrogen burning occupies $\simeq90\%$ of the lifetime.
Assuming the linear $v_{\mathrm c}(t)$ trend extends through the remaining
nuclear stages, we combine the late‑MS $v_{\mathrm c}$ with the measured
convective‑core masses $M_{\rm c}$ to estimate the Kerr parameter of the direct-collapse black hole (DCBH).  Table~\ref{tab:bh_spin} gives results for two limits:
(i) the DCBH accretes only the envelope above the CO core
(\,$M_{\rm BH}=M_{\rm c}$\,) and (ii) the entire star collapses
(\,$M_{\rm BH}=M_\star$\,).
The most extreme model,  $9000\,M_\odot$ with a core velocity
$v_{\mathrm c}\simeq120\;\mathrm{km\,s^{-1}}$, still reaches only a spin parameter 
$a_*^{\rm(core)}\!\approx\!0.05$, far below the Thorne limit. Full collapse
reduces the spin to $a_*^{\rm(full)}\!\approx\!0.02$.
If future simulations yield higher spins, the excess angular momentum must be gained after core‑H exhaustion—e.g. by renewed shear during He burning, secular contraction in the final $\sim10^4$ yr, or late‑stage fallback.

The spin estimate conserves angular momentum, not linear speed:
$J_{\rm c}=kM_{\rm c}R_{\rm c}^{2}(v_{\rm c}/R_{\rm c})$ is carried unchanged into the collapse, while the physical radius shrinks by five to six orders of magnitude. The Kerr parameter $a_*=cJ_{\rm c}/(G M_{\rm BH}^{2})$ therefore remains fixed even though the horizon rotates at $v\!=\!a_*c$. Any higher IMBH spin must be accumulated \emph{after} core-H exhaustion,
for example through shear generated in core-He burning, secular
contraction during the final $10^{4}$ yr, or late fallback.

\begin{table}
\centering
\caption{Dimensionless spin parameters $a_*$ at core–H exhaustion.}
\label{tab:bh_spin}
\begin{tabular}{cccccc}
\hline\hline
$M_\star$ & $v/v_{\rm crit}$ & $M_{\rm c}$ & $a_*^{\rm(core)}$ & $a_*^{\rm(full)}$\\
($M_\odot$) & & ($M_\odot$) &  &  \\
\hline
1000 & 0.01 & 400  & 0.09 & 0.015 \\
     & 0.05 & 510  & 0.10 & 0.026 \\
     & 0.10 & 560  & 0.12 & 0.037 \\[2pt]
\hline
5000 & 0.01 & 2700 & 0.035 & 0.010 \\
     & 0.05 & 2900 & 0.042 & 0.014 \\
     & 0.10 & 3000 & 0.050 & 0.018 \\[2pt]
\hline
9000 & 0.01 & 5100 & 0.030 & 0.009 \\
     & 0.05 & 5600 & 0.034 & 0.013 \\
     & 0.10 & 5760 & 0.052 & 0.021 \\
\hline
\end{tabular}
\end{table}

\section{Discussion}\label{Sec:Discussion}

\subsection{Comparison with non‑rotating models}\label{Sec:DiscCompare}
\begin{figure}[!h]
  \centering
  \includegraphics[width=9cm]{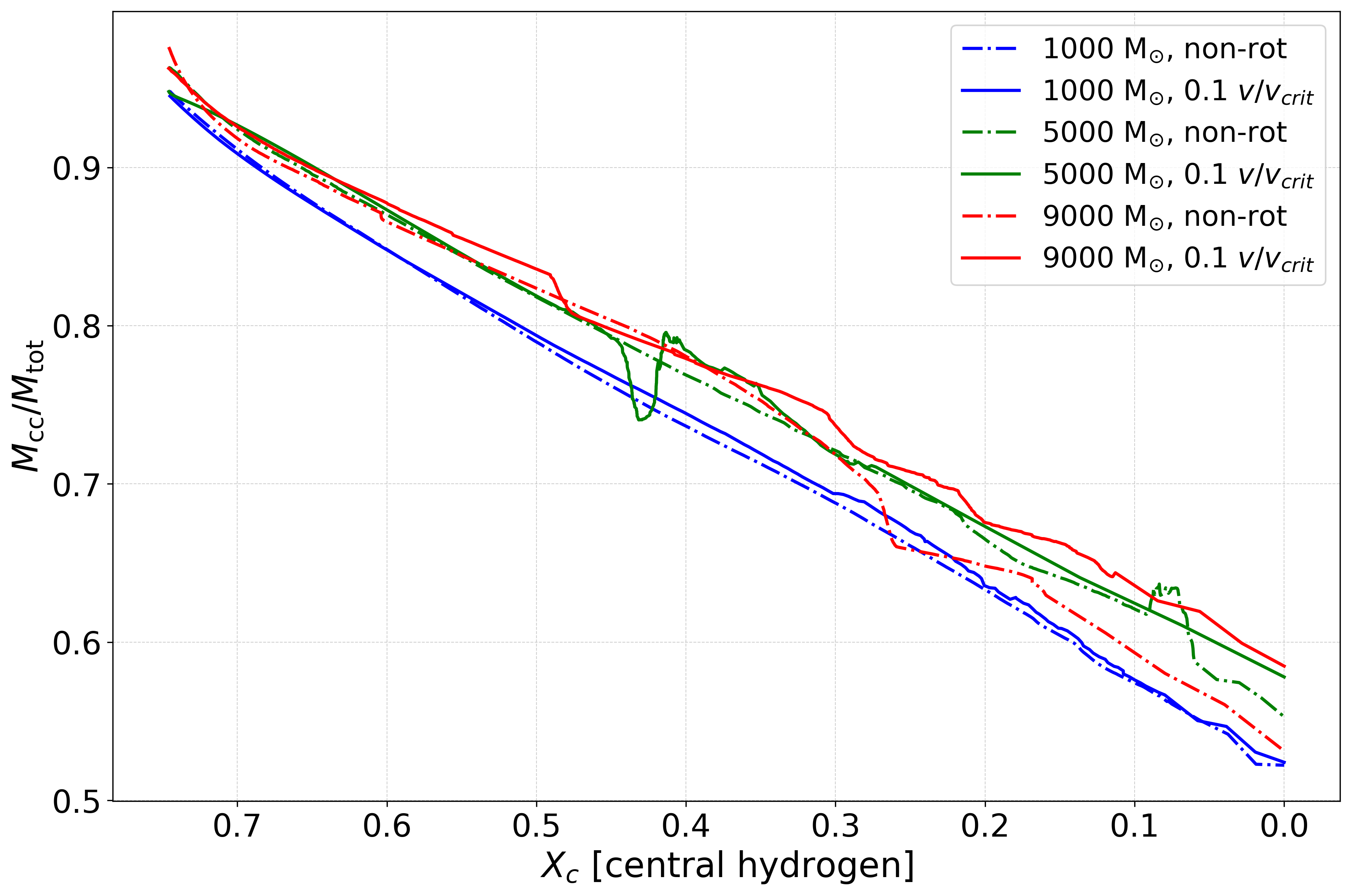}
  \caption{\(M_{\rm cc}/M_{\rm tot}\) versus central hydrogen \(X_{\rm c}\) for non‑rotating (dash‑dot) and \(v/v_{\rm crit}=0.1\) (solid) models at 1000 (blue), 5000 (green) and 9000 (red) \Ms.}
  \label{fig:Mcc_vs_Xc}
\end{figure}

We compare our rotating SMS tracks, specifically the $v/v_{crit}=0.1$ models, with the non-rotating Pop III models of \citet{Nandal2025} at the same \(X_c\) (Fig.~\ref{fig:Mcc_vs_Xc}). The convective core mass fraction ($M_{\rm cc}$/$M_{ \rm tot}$) reveals how rotation alters core growth against steep radiative and compositional gradients. At 1000 \Ms, rotation increases the mid‐MS core fraction to $M_{\rm cc}/M_{\rm tot}=0.780$ compared to 0.744 without rotation (+4.6\%), and the end‐MS core reaches 0.521 versus 0.512 (+1.8\%). During the early MS, meridional circulation ($U_2$) dredges hydrogen across the $\mu$‐gradient, sustained by shear instabilities ($D_{\rm shear}$), which enlarges the convective core. As the envelope inflates and shear weakens late in the MS, core growth under rotation is partly counteracted, limiting the end‐MS gain.

At 5000 \Ms, rotating models show a mid‐MS $M_{cc}/M_{\rm tot}=0.741$ versus 0.738 non‐rot (+0.4\%), but by end‐MS the core grows from 0.555 to 0.570 (+2.7\%). The higher luminosity steepens the radiative gradient ($\nabla_{\rm rad}$), yet rotation amplifies vertical shear mixing ($D_{\rm shear}\sim10^{12}\,\rm cm^2\,s^{-1}$) and effective diffusion ($D_{\rm eff}\sim10^9\,\rm cm^2\,s^{-1}$), allowing more hydrogen to penetrate the core throughout the MS. Reduced envelope density late in evolution further enhances this mixing.

At 9000 \Ms, the rotating mid‐MS core fraction (0.736 at $X_c\approx0.113$) falls below the non‐rotating 0.774 (–4.9\%) because near‐Eddington envelopes ($\Gamma_{\rm edd}\gtrsim0.96$) initially suppress meridional flows. As the star evolves, centrifugal support reduces the effective gravity ($g_{\rm eff}$), meridional velocities recover, and by H exhaustion the rotating core reaches 0.585 versus 0.541 non‐rot (+8.1\%). This late‐MS core boost highlights how rotation can overcome radiative damping to sustain mixing in the most massive SMS.

These end‑MS core enhancements translate into MS lifetime extensions of $\sim$ 7\%, 14\% and 20\% at 1000, 5000 and 9000 \Ms, respectively.  At matched \(X_{\rm c}\), rotating models are cooler by \(\Delta\log T_{\rm eff}\approx-0.76\), \(-0.04\) and \(\sim0\), and luminosities shift by \(\Delta\log L\approx+0.04\), \(-0.03\) and \(+0.002\).  
Eddington factors at these points decrease from 0.894 to 0.864, 0.977 to 0.961, and 0.993 to 0.965, respectively, affecting the onset of mechanical mass loss.

\subsection{Comparison with other works}\label{Sec:CompareOther}

\citet{Lee2016} evolved one‑dimensional, accreting Pop III protostars at \(\dot M\simeq10^{-3}\,M_\odot\,\mathrm{yr^{-1}}\) and showed that once \(M\approx5\)–\(7\,M_\odot\), envelope inflation drives \(\Gamma\to1\) so that $g_{\rm eff}(1-\Gamma)=0$ is reached at \(v/v_{\rm crit}\ll1\), halting further accretion by enforcing the \(\Omega\Gamma\)–limit and capping growth at \(\sim20\)–\(40\,M_\odot\) while keeping radii compact (\(R\lesssim50\,R_\odot\)).  They interpret this as a complete termination of protostellar evolution beyond that mass.  Our supermassive models begin at \(10^3\)–\(10^4\,M_\odot\) with initial \(v/v_{\rm crit}=0.01\)–0.10, well below the \(\sim0.7\) threshold for \(\Omega\Gamma\)–driven break‑up.  Although our SMSs cross the \(\Omega\Gamma\)–limit, they shed \(\lesssim2\%\) of their mass yet continue core H burning to \(X_c\to0\), demonstrating that this combined radiative‐centrifugal limit regulates but does not terminate stellar evolution once formation is complete.

\cite{Yoon2012} computed a grid of rotating Pop III stars from 10 -  1000 \Ms, including magnetic torques and rotational mixing.  They find that chemically homogeneous evolution (CHE) occurs for $13 \le M \le 84$ \Ms\ at $v/v_{\rm crit} \gtrsim$ 0.4 and ceases above $\sim$ 190 \Ms\ when envelope inflation and angular‐momentum loss suppress mixing.  Although the CHE threshold parallels the \(\Omega\Gamma\)–limit onset at moderate masses, our SMS models at \(v/v_{\rm crit}\le0.10\) remain outside the CHE regime but nonetheless exhibit significant core‐mass enhancements via meridional circulation and shear mixing (Sect.~\ref{Sec:DiscCompare}), illustrating that sub‑CHE rotation can still measurably alter supermassive stellar structure.

With the formalism developed by \citet{OGlimit2000}, \citet{Tsiatsiou2024} recently used GENEC to compute fast-rotating ($v/v_{crit}=0.7$) Pop~III models in the 9--120 \Ms\ range. They found that despite approaching the $\Omega\Gamma$ limit, their models lost $\le 1\%$ of their mass. Our SMS models differ in that they are much more massive and luminous, but are rotating much more slowly ($v/v_{crit} \le 0.1$). The key distinction is that the high Eddington factors in our models ($\Gamma_{Edd} \ge 0.96$) drastically lower the true critical velocity ($v_{crit,2}$), as seen in the bottom panel of Figure~\ref{fig:OOc_vs_Gamma}. This means that even modest rotation is sufficient to push the star against the $\Omega\Gamma$ limit. Like Tsiatsiou et al. (2024), we find that this results in regulation via moderate mass loss ($\le 2\%$) rather than catastrophic disruption, confirming that this physical process operates across a wide range of stellar masses. 

\citet{Haemmerle2018} followed accretion and rotation in Pop III protostars up to supermassive scales and showed that the $\Omega\Gamma$ limit requires the accreted angular momentum to be \(\lesssim1\%\) of the Keplerian value. This yields surface velocities \(v/v_{\rm crit}\lesssim0.1\)–0.2 to avoid premature break‑up and allow growth to \(10^5\,M_\odot\).  The constraint here enforces slow rotation and negligible centrifugal deformation during accretion.  Our rotating SMS models lie precisely within the slow rotation window identified by \citet{Haemmerle2018} - and although they cross the $\Omega\Gamma$ limit during core H burning, they shed \(\lesssim2\%\) of their mass and exhaust hydrogen to \(X_c\to0\), demonstrating that the same $\Omega\Gamma$ regulation applies post‑assembly without halting stellar evolution.  

Finally, it is worth considering the effect of metallicity. Our models are strictly for metal-free (Pop~III) stars. The formation of SMSs may also be possible in very low-metallicity gas \citep{Chon2020}. The presence of even trace amounts of metals would significantly enhance mass-loss rates from line-driven winds. As shown in recent work \citep{Sibony2025}, this would alter the angular momentum evolution and could change the final fate and black hole remnant properties, representing an important avenue for future research.

\section{Conclusion}\label{Sec:Conclusion}

We have performed the first simulations of rotating supermassive Pop III star evolution with GENEC. These models examine the interplay of rotation-induced mixing, angular momentum transport, and the $\Omega\Gamma$ limit to advance our understanding of 1000 - 10000 \Ms\ primordial stellar evolution.  Our main conclusions can be summarized as follows:
\begin{itemize}
\item Rotation notably extends the main-sequence lifetime of supermassive Pop III stars by approximately 7\% at 1000 $M_\odot$, 14\% at 5000 $M_\odot$, and 20\% at 9000 $M_\odot$ relative to non-rotating models. This is primarily due to enhanced chemical mixing that enlarges the convective core by up to 8.1\% in the most massive cases.
\item The $\Omega\Gamma$ limit significantly constrains stellar rotation, triggering mechanical mass loss once surface rotation approaches approximately 70\% of the critical velocity. However, total mass loss (not accounting for pulsational mass loss) remains modest ($\lesssim2\%$), suggesting regulation rather than cessation of stellar evolution.
\item Internal angular momentum transport via meridional circulation and shear instabilities efficiently redistributes angular momentum, substantially decelerating core rotation. Predicted dimensionless spin parameters for resulting DCBHs remain low ($a_* \le 0.05$ for core collapse scenarios). This implies that rotating SMSs are natural progenitors of slowly rotating black hole seeds. If rapidly spinning seeds are required by observations, they must acquire additional angular momentum through post-main-sequence processes, such as late-stage accretion or fallback.
\item Chemical mixing in rotating SMSs is influenced by convective mixing within core and intermediate shells, meridional circulation-driven effective mixing ($D_{\rm eff}\sim10^9$ cm$^2$ s$^{-1}$), and shear-induced vertical mixing ($D_{\rm shear}\sim10^{12}$ cm$^2$ s$^{-1}$), each critically shaping chemical gradients and core evolution. However, as the structure of the stars is primarily convetive, $D_{\rm conv}$ is the dominant mixing coefficient for chemical transport throughout the main sequence. 

\end{itemize}

In comparison with previous non-rotating Pop III models, rotation consistently produces larger convective cores and longer stellar lifetimes, a result consistent with lower-mass massive stars. Unlike earlier lower-mass Pop III studies encountering the $\Omega\Gamma$ limit, our supermassive models demonstrate only moderate mass loss and continued stable evolution through core hydrogen exhaustion. These findings align with recent theoretical insights that rotation significantly impacts supermassive star structure without necessarily triggering catastrophic mass loss.

Future investigations will extend these rotating stellar models into higher rotation rates and advanced stages of nuclear burning \citep{Griffiths2025}, explicitly calculating nucleosynthetic yields essential for interpreting high-redshift abundance observations. Additionally, implementing metallicity-dependent mass-loss prescriptions will enhance realism and applicability to early-universe conditions. Inclusion of magnetic instabilities \citep{Spruit2002, Fuller2019, Griffiths2022} will further refine angular momentum transport modeling, which might be particularly important in light of the observations made for low mass stars for which asteroseismic constraints are available \citep{Aerts2019, Buldgen2023}. Finally, exploring the combined effects of rotation and accretion from the pre-main sequence phase will provide comprehensive evolutionary pathways crucial for understanding the formation and ultimate fate of the earliest supermassive stars.

\begin{acknowledgements}
D.N. would like to thank Prof. André Maeder for all the valuable discussions and inspiration for this work. GB acknowledges fundings from the Fonds National de la Recherche Scientifique (FNRS) as a postdoctoral researcher. JR acknowledges support from the Royal Society and Research Ireland through the University Research Fellow programme under grant number URF$\backslash$R1$\backslash$191132 and acknowledges support from the Research Ireland Laureate programme under grant number IRCLA/2022/1165. T.E.W. acknowledges the support of the Canadian Space Agency (CSA) [{\bf 23EXPROSS2}] and the support of the Natural Sciences and Engineering Research Council of Canada (NSERC).
\end{acknowledgements}

\bibliographystyle{aa}
\bibliography{biblio}

\begin{thebibliography}{84}
\expandafter\ifx\csname natexlab\endcsname\relax\def\natexlab#1{#1}\fi

\bibitem[{{Aerts} {et~al.}(2019){Aerts}, {Mathis}, \& {Rogers}}]{Aerts2019}
{Aerts}, C., {Mathis}, S., \& {Rogers}, T.~M. 2019, \araa, 57, 35

\bibitem[{{Ba{\~n}ados} {et~al.}(2018){Ba{\~n}ados}, {Venemans}, {Mazzucchelli}, {Farina}, {Walter}, {Wang}, {Decarli}, {Stern}, {Fan}, {Davies}, {Hennawi}, {Simcoe}, {Turner}, {Rix}, {Yang}, {Kelson}, {Rudie}, \& {Winters}}]{Banados2018}
{Ba{\~n}ados}, E., {Venemans}, B.~P., {Mazzucchelli}, C., {et~al.} 2018, \nat, 553, 473

\bibitem[{{Banik} {et~al.}(2019){Banik}, {Tan}, \& {Monaco}}]{Banik2019}
{Banik}, N., {Tan}, J.~C., \& {Monaco}, P. 2019, \mnras, 483, 3592

\bibitem[{{Barausse}(2012)}]{Barausse2012}
{Barausse}, E. 2012, \mnras, 423, 2533

\bibitem[{{Berti} \& {Volonteri}(2008)}]{Berti2008}
{Berti}, E. \& {Volonteri}, M. 2008, \apj, 684, 822

\bibitem[{{Brott} {et~al.}(2011){Brott}, {de Mink}, {Cantiello}, {Langer}, {de Koter}, {Evans}, {Hunter}, {Trundle}, \& {Vink}}]{Brott2011}
{Brott}, I., {de Mink}, S.~E., {Cantiello}, M., {et~al.} 2011, \aap, 530, A115

\bibitem[{{Buldgen} \& {Eggenberger}(2023)}]{Buldgen2023}
{Buldgen}, G. \& {Eggenberger}, P. 2023, in The Sixteenth Marcel Grossmann Meeting. On Recent Developments in Theoretical and Experimental General Relativity, Astrophysics, and Relativistic Field Theories, ed. R.~{Ruffino} \& G.~{Vereshchagin}, 2848--2864

\bibitem[{{Bunker} {et~al.}(2023){Bunker}, {Saxena}, {Cameron}, {Willott}, {Curtis-Lake}, {Jakobsen}, {Carniani}, {Smit}, {Maiolino}, {Witstok}, {Curti}, {D'Eugenio}, {Jones}, {Ferruit}, {Arribas}, {Charlot}, {Chevallard}, {Giardino}, {de Graaff}, {Looser}, {L{\"u}tzgendorf}, {Maseda}, {Rawle}, {Rix}, {Del Pino}, {Alberts}, {Egami}, {Eisenstein}, {Endsley}, {Hainline}, {Hausen}, {Johnson}, {Rieke}, {Rieke}, {Robertson}, {Shivaei}, {Stark}, {Sun}, {Tacchella}, {Tang}, {Williams}, {Willmer}, {Baker}, {Baum}, {Bhatawdekar}, {Bowler}, {Boyett}, {Chen}, {Circosta}, {Helton}, {Ji}, {Kumari}, {Lyu}, {Nelson}, {Parlanti}, {Perna}, {Sandles}, {Scholtz}, {Suess}, {Topping}, {{\"U}bler}, {Wallace}, \& {Whitler}}]{Bunker2023}
{Bunker}, A.~J., {Saxena}, A., {Cameron}, A.~J., {et~al.} 2023, \aap, 677, A88

\bibitem[{{Cameron} {et~al.}(2023){Cameron}, {Katz}, {Rey}, \& {Saxena}}]{Cameron2023}
{Cameron}, A.~J., {Katz}, H., {Rey}, M.~P., \& {Saxena}, A. 2023, \mnras

\bibitem[{{Chon} \& {Omukai}(2020)}]{Chon2020}
{Chon}, S. \& {Omukai}, K. 2020, \mnras, 494, 2851

\bibitem[{{Eggenberger} {et~al.}(2008){Eggenberger}, {Meynet}, {Maeder}, {Hirschi}, {Charbonnel}, {Talon}, \& {Ekstr{\"o}m}}]{Eggenberger2008}
{Eggenberger}, P., {Meynet}, G., {Maeder}, A., {et~al.} 2008, \apss, 316, 43

\bibitem[{{Ekstr{\"o}m} {et~al.}(2012){Ekstr{\"o}m}, {Georgy}, {Eggenberger}, {Meynet}, {Mowlavi}, {Wyttenbach}, {Granada}, {Decressin}, {Hirschi}, {Frischknecht}, {Charbonnel}, \& {Maeder}}]{Ekstrom2012}
{Ekstr{\"o}m}, S., {Georgy}, C., {Eggenberger}, P., {et~al.} 2012, \aap, 537, A146

\bibitem[{{Ekstr{\"o}m} {et~al.}(2008){Ekstr{\"o}m}, {Meynet}, {Chiappini}, {Hirschi}, \& {Maeder}}]{Ekstrom2008}
{Ekstr{\"o}m}, S., {Meynet}, G., {Chiappini}, C., {Hirschi}, R., \& {Maeder}, A. 2008, \aap, 489, 685

\bibitem[{{Fryer} \& {Heger}(2011)}]{Fryer2011}
{Fryer}, C.~L. \& {Heger}, A. 2011, Astronomische Nachrichten, 332, 408

\bibitem[{{Fuller} {et~al.}(2019){Fuller}, {Piro}, \& {Jermyn}}]{Fuller2019}
{Fuller}, J., {Piro}, A.~L., \& {Jermyn}, A.~S. 2019, \mnras, 485, 3661

\bibitem[{{Glatzel}(1998)}]{Glatzel1998}
{Glatzel}, W. 1998, \aap, 339, L5

\bibitem[{{Griffiths} {et~al.}(2025){Griffiths}, {Aloy}, {Hirschi}, {Reichert}, {Obergaulinger}, {Whitehead}, {Martinet}, {Sciarini}, {Ekstr{\"o}m}, \& {Meynet}}]{Griffiths2025}
{Griffiths}, A., {Aloy}, M.-{\'A}., {Hirschi}, R., {et~al.} 2025, \aap, 693, A93

\bibitem[{{Griffiths} {et~al.}(2022){Griffiths}, {Eggenberger}, {Meynet}, {Moyano}, \& {Aloy}}]{Griffiths2022}
{Griffiths}, A., {Eggenberger}, P., {Meynet}, G., {Moyano}, F., \& {Aloy}, M.-{\'A}. 2022, \aap, 665, A147

\bibitem[{{Haemmerl{\'e}}(2021)}]{Lionel2021}
{Haemmerl{\'e}}, L. 2021, \aap, 650, A204

\bibitem[{{Haemmerl{\'e}} {et~al.}(2018){Haemmerl{\'e}}, {Woods}, {Klessen}, {Heger}, \& {Whalen}}]{Haemmerle2018}
{Haemmerl{\'e}}, L., {Woods}, T.~E., {Klessen}, R.~S., {Heger}, A., \& {Whalen}, D.~J. 2018, \apjl, 853, L3

\bibitem[{{Heger} {et~al.}(2000){Heger}, {Langer}, \& {Woosley}}]{Heger2000}
{Heger}, A., {Langer}, N., \& {Woosley}, S.~E. 2000, \apj, 528, 368

\bibitem[{{Herrington} {et~al.}(2023){Herrington}, {Whalen}, \& {Woods}}]{herr23a}
{Herrington}, N.~P., {Whalen}, D.~J., \& {Woods}, T.~E. 2023, \mnras, 521, 463

\bibitem[{{Hirano} \& {Bromm}(2018)}]{Hirano2018}
{Hirano}, S. \& {Bromm}, V. 2018, \mnras, 476, 3964

\bibitem[{{Inayoshi} {et~al.}(2020){Inayoshi}, {Visbal}, \& {Haiman}}]{Inayoshi2020}
{Inayoshi}, K., {Visbal}, E., \& {Haiman}, Z. 2020, \araa, 58, 27

\bibitem[{{Ishiyama} \& {Hirano}(2025)}]{Ishiyama2025}
{Ishiyama}, T. \& {Hirano}, S. 2025, arXiv e-prints, arXiv:2501.17540

\bibitem[{{Isobe} {et~al.}(2023){Isobe}, {Ouchi}, {Tominaga}, {Watanabe}, {Nakajima}, {Umeda}, {Yajima}, {Harikane}, {Fukushima}, {Xu}, {Ono}, \& {Zhang}}]{Isobe2023}
{Isobe}, Y., {Ouchi}, M., {Tominaga}, N., {et~al.} 2023, \apj, 959, 100

\bibitem[{{Kimura} {et~al.}(2023){Kimura}, {Hosokawa}, {Sugimura}, \& {Fukushima}}]{Kimura2023}
{Kimura}, K., {Hosokawa}, T., {Sugimura}, K., \& {Fukushima}, H. 2023, \apj, 950, 184

\bibitem[{{Kiyuna} {et~al.}(2023){Kiyuna}, {Hosokawa}, \& {Chon}}]{Kiyuna2023}
{Kiyuna}, M., {Hosokawa}, T., \& {Chon}, S. 2023, \mnras, 523, 1496

\bibitem[{{Kiyuna} {et~al.}(2024){Kiyuna}, {Hosokawa}, \& {Chon}}]{Kiyuna2024}
{Kiyuna}, M., {Hosokawa}, T., \& {Chon}, S. 2024, \mnras, 534, 3916

\bibitem[{{Kudritzki}(2002)}]{Kudritzki2002}
{Kudritzki}, R.~P. 2002, \apj, 577, 389

\bibitem[{{Langer}(1998)}]{Langer1998}
{Langer}, N. 1998, \aap, 329, 551

\bibitem[{{Larson} {et~al.}(2023){Larson}, {Finkelstein}, {Kocevski}, {Hutchison}, {Trump}, {Arrabal Haro}, {Bromm}, {Cleri}, {Dickinson}, {Fujimoto}, {Kartaltepe}, {Koekemoer}, {Papovich}, {Pirzkal}, {Tacchella}, {Zavala}, {Bagley}, {Behroozi}, {Champagne}, {Cole}, {Jung}, {Morales}, {Yang}, {Zhang}, {Zitrin}, {Amor{\'\i}n}, {Burgarella}, {Casey}, {Ch{\'a}vez Ortiz}, {Cox}, {Chworowsky}, {Fontana}, {Gawiser}, {Grazian}, {Grogin}, {Harish}, {Hathi}, {Hirschmann}, {Holwerda}, {Juneau}, {Leung}, {Lucas}, {McGrath}, {P{\'e}rez-Gonz{\'a}lez}, {Rigby}, {Seill{\'e}}, {Simons}, {de La Vega}, {Weiner}, {Wilkins}, {Yung}, \& {Ceers Team}}]{Larson2023}
{Larson}, R.~L., {Finkelstein}, S.~L., {Kocevski}, D.~D., {et~al.} 2023, \apjl, 953, L29

\bibitem[{{Latif} {et~al.}(2021){Latif}, {Khochfar}, {Schleicher}, \& {Whalen}}]{latif21a}
{Latif}, M.~A., {Khochfar}, S., {Schleicher}, D., \& {Whalen}, D.~J. 2021, \mnras, 508, 1756

\bibitem[{{Latif} {et~al.}(2014){Latif}, {Schleicher}, {Bovino}, {Grassi}, \& {Spaans}}]{Latif_2014b}
{Latif}, M.~A., {Schleicher}, D.~R.~G., {Bovino}, S., {Grassi}, T., \& {Spaans}, M. 2014, \apj, 792, 78

\bibitem[{{Latif} {et~al.}(2016){Latif}, {Schleicher}, \& {Hartwig}}]{Latif2016}
{Latif}, M.~A., {Schleicher}, D.~R.~G., \& {Hartwig}, T. 2016, \mnras, 458, 233

\bibitem[{{Latif} {et~al.}(2022{\natexlab{a}}){Latif}, {Whalen}, {Khochfar}, {Herrington}, \& {Woods}}]{Latif_2022}
{Latif}, M.~A., {Whalen}, D.~J., {Khochfar}, S., {Herrington}, N.~P., \& {Woods}, T.~E. 2022{\natexlab{a}}, \nat, 607, 48

\bibitem[{{Latif} {et~al.}(2022{\natexlab{b}}){Latif}, {Whalen}, {Khochfar}, {Herrington}, \& {Woods}}]{Latif2022}
{Latif}, M.~A., {Whalen}, D.~J., {Khochfar}, S., {Herrington}, N.~P., \& {Woods}, T.~E. 2022{\natexlab{b}}, \nat, 607, 48

\bibitem[{{Lee} \& {Yoon}(2016)}]{Lee2016}
{Lee}, H. \& {Yoon}, S.-C. 2016, \apj, 820, 135

\bibitem[{{Limongi} \& {Chieffi}(2018)}]{Limongi2018}
{Limongi}, M. \& {Chieffi}, A. 2018, \apjs, 237, 13

\bibitem[{{Liu} {et~al.}(2024){Liu}, {Gurian}, {Inayoshi}, {Hirano}, {Hosokawa}, {Bromm}, \& {Yoshida}}]{Liu2024}
{Liu}, B., {Gurian}, J., {Inayoshi}, K., {et~al.} 2024, \mnras, 534, 290

\bibitem[{{Liu} {et~al.}(2025){Liu}, {Sibony}, {Meynet}, \& {Bromm}}]{Liu2025}
{Liu}, B., {Sibony}, Y., {Meynet}, G., \& {Bromm}, V. 2025, \apjl, 980, L30

\bibitem[{{Maeder}(1997)}]{Maeder1997}
{Maeder}, A. 1997, \aap, 321, 134

\bibitem[{{Maeder} \& {Meynet}(2000)}]{OGlimit2000}
{Maeder}, A. \& {Meynet}, G. 2000, \aap, 361, 159

\bibitem[{{Marques-Chaves} {et~al.}(2024){Marques-Chaves}, {Schaerer}, {Kuruvanthodi}, {Korber}, {Prantzos}, {Charbonnel}, {Weibel}, {Izotov}, {Messa}, {Brammer}, {Dessauges-Zavadsky}, \& {Oesch}}]{Marques2023}
{Marques-Chaves}, R., {Schaerer}, D., {Kuruvanthodi}, A., {et~al.} 2024, \aap, 681, A30

\bibitem[{{Mortlock} {et~al.}(2011){Mortlock}, {Warren}, {Venemans}, {Patel}, {Hewett}, {McMahon}, {Simpson}, {Theuns}, {Gonz{\'a}les-Solares}, {Adamson}, {Dye}, {Hambly}, {Hirst}, {Irwin}, {Kuiper}, {Lawrence}, \& {R{\"o}ttgering}}]{Mortlock2011}
{Mortlock}, D.~J., {Warren}, S.~J., {Venemans}, B.~P., {et~al.} 2011, \nat, 474, 616

\bibitem[{{Murphy} {et~al.}(2021){Murphy}, {Groh}, {Ekstr{\"o}m}, {Meynet}, {Pezzotti}, {Georgy}, {Choplin}, {Eggenberger}, {Farrell}, {Haemmerl{\'e}}, {Hirschi}, {Maeder}, \& {Martinet}}]{Murphy2021a}
{Murphy}, L.~J., {Groh}, J.~H., {Ekstr{\"o}m}, S., {et~al.} 2021, \mnras, 501, 2745

\bibitem[{{Nagele} \& {Umeda}(2023)}]{Nagele2023}
{Nagele}, C. \& {Umeda}, H. 2023, \apjl, 949, L16

\bibitem[{{Nakajima} {et~al.}(2023){Nakajima}, {Ouchi}, {Isobe}, {Harikane}, {Zhang}, {Ono}, {Umeda}, \& {Oguri}}]{Nakajima2023}
{Nakajima}, K., {Ouchi}, M., {Isobe}, Y., {et~al.} 2023, \apjs, 269, 33

\bibitem[{{Nandal} {et~al.}(2024{\natexlab{a}}){Nandal}, {Farrell}, {Buldgen}, {Meynet}, \& {Ekstr{\"o}m}}]{Nandal2024d}
{Nandal}, D., {Farrell}, E., {Buldgen}, G., {Meynet}, G., \& {Ekstr{\"o}m}, S. 2024{\natexlab{a}}, \aap, 685, A159

\bibitem[{{Nandal} {et~al.}(2024{\natexlab{b}}){Nandal}, {Meynet}, {Ekstr{\"o}m}, {Moyano}, {Eggenberger}, {Choplin}, {Georgy}, {Farrell}, \& {Maeder}}]{Nandal2024b}
{Nandal}, D., {Meynet}, G., {Ekstr{\"o}m}, S., {et~al.} 2024{\natexlab{b}}, \aap, 684, A169

\bibitem[{{Nandal} {et~al.}(2023){Nandal}, {Regan}, {Woods}, {Farrell}, {Ekstr{\"o}m}, \& {Meynet}}]{Nandal2023}
{Nandal}, D., {Regan}, J.~A., {Woods}, T.~E., {et~al.} 2023, \aap, 677, A155

\bibitem[{{Nandal} {et~al.}(2024{\natexlab{c}}){Nandal}, {Regan}, {Woods}, {Farrell}, {Ekstr{\"o}m}, \& {Meynet}}]{Nandal2024}
{Nandal}, D., {Regan}, J.~A., {Woods}, T.~E., {et~al.} 2024{\natexlab{c}}, \aap, 683, A156

\bibitem[{{Nandal} {et~al.}(2024{\natexlab{d}}){Nandal}, {Sibony}, \& {Tsiatsiou}}]{Nandal2024e}
{Nandal}, D., {Sibony}, Y., \& {Tsiatsiou}, S. 2024{\natexlab{d}}, \aap, 688, A142

\bibitem[{{Nandal} {et~al.}(2025{\natexlab{a}}){Nandal}, {Topalakis}, {Tan}, {Sergienko}, {Pauchett}, \& {Petkova}}]{Nandal2025c}
{Nandal}, D., {Topalakis}, K., {Tan}, J.~C., {et~al.} 2025{\natexlab{a}}, arXiv e-prints, arXiv:2507.00870

\bibitem[{{Nandal} {et~al.}(2025{\natexlab{b}}){Nandal}, {Whalen}, {Latif}, \& {Heger}}]{Nandal2025}
{Nandal}, D., {Whalen}, D.~J., {Latif}, M.~A., \& {Heger}, A. 2025{\natexlab{b}}, arXiv e-prints, arXiv:2502.04435

\bibitem[{{Owocki} {et~al.}(2004){Owocki}, {Gayley}, \& {Shaviv}}]{Owocki2004}
{Owocki}, S.~P., {Gayley}, K.~G., \& {Shaviv}, N.~J. 2004, \apj, 616, 525

\bibitem[{{Patrick} {et~al.}(2023){Patrick}, {Whalen}, {Latif}, \& {Elford}}]{pat23a}
{Patrick}, S.~J., {Whalen}, D.~J., {Latif}, M.~A., \& {Elford}, J.~S. 2023, \mnras, 522, 3795

\bibitem[{{Regan}(2023)}]{Regan_2022}
{Regan}, J. 2023, The Open Journal of Astrophysics, 6, 12

\bibitem[{{Regan} {et~al.}(2017){Regan}, {Visbal}, {Wise}, , {Haiman}, {Johansson}, \& {Bryan}}]{Regan_2017}
{Regan}, J.~A., {Visbal}, E., {Wise}, J.~H., {et~al.} 2017, Nature Astronomy, 1, 0075

\bibitem[{{Regan} {et~al.}(2020{\natexlab{a}}){Regan}, {Wise}, {O'Shea}, \& {Norman}}]{Regan_2020}
{Regan}, J.~A., {Wise}, J.~H., {O'Shea}, B.~W., \& {Norman}, M.~L. 2020{\natexlab{a}}, \mnras, 492, 3021

\bibitem[{{Regan} {et~al.}(2020{\natexlab{b}}){Regan}, {Wise}, {Woods}, {Downes}, {O'Shea}, \& {Norman}}]{Regan_2020b}
{Regan}, J.~A., {Wise}, J.~H., {Woods}, T.~E., {et~al.} 2020{\natexlab{b}}, The Open Journal of Astrophysics, 3, 15

\bibitem[{{Sanyal} {et~al.}(2015){Sanyal}, {Grassitelli}, {Langer}, \& {Bestenlehner}}]{Sanyal2015}
{Sanyal}, D., {Grassitelli}, L., {Langer}, N., \& {Bestenlehner}, J.~M. 2015, \aap, 580, A20

\bibitem[{{Schauer} {et~al.}(2017{\natexlab{a}}){Schauer}, {Regan}, {Glover}, \& {Klessen}}]{Schauer_2017}
{Schauer}, A.~T.~P., {Regan}, J., {Glover}, S.~C.~O., \& {Klessen}, R.~S. 2017{\natexlab{a}}, \mnras, 471, 4878

\bibitem[{{Schauer} {et~al.}(2017{\natexlab{b}}){Schauer}, {Regan}, {Glover}, \& {Klessen}}]{Schauer2017}
{Schauer}, A. T.~P., {Regan}, J., {Glover}, S. C.~O., \& {Klessen}, R.~S. 2017{\natexlab{b}}, \mnras, 471, 4878

\bibitem[{{Sesana} {et~al.}(2009){Sesana}, {Volonteri}, \& {Haardt}}]{Sesana2009}
{Sesana}, A., {Volonteri}, M., \& {Haardt}, F. 2009, Classical and Quantum Gravity, 26, 094033

\bibitem[{{Sibony} {et~al.}(2024){Sibony}, {Shepherd}, {Yusof}, {Hirschi}, {Chambers}, {Tsiatsiou}, {Nandal}, {Sciarini}, {Moyano}, {B{\'e}trisey}, {Buldgen}, {Georgy}, {Ekstr{\"o}m}, {Eggenberger}, \& {Meynet}}]{Sibony2025}
{Sibony}, Y., {Shepherd}, K.~G., {Yusof}, N., {et~al.} 2024, \aap, 690, A91

\bibitem[{{Singh} {et~al.}(2023){Singh}, {Monaco}, \& {Tan}}]{Singh2023}
{Singh}, J., {Monaco}, P., \& {Tan}, J.~C. 2023, \mnras, 525, 969

\bibitem[{{Smidt} {et~al.}(2018){Smidt}, {Whalen}, {Johnson}, {Surace}, \& {Li}}]{Smidt18}
{Smidt}, J., {Whalen}, D.~J., {Johnson}, J.~L., {Surace}, M., \& {Li}, H. 2018, \apj, 865, 126

\bibitem[{{Spolyar} {et~al.}(2008){Spolyar}, {Freese}, \& {Gondolo}}]{Spolyar2008}
{Spolyar}, D., {Freese}, K., \& {Gondolo}, P. 2008, \prl, 100, 051101

\bibitem[{{Spruit}(2002)}]{Spruit2002}
{Spruit}, H.~C. 2002, \aap, 381, 923

\bibitem[{{Tan} {et~al.}(2024){Tan}, {Singh}, {Cammelli}, {Sanati}, {Petkova}, {Nandal}, \& {Monaco}}]{Tan2024}
{Tan}, J.~C., {Singh}, J., {Cammelli}, V., {et~al.} 2024, arXiv e-prints, arXiv:2412.01828

\bibitem[{{Tsiatsiou} {et~al.}(2024){Tsiatsiou}, {Sibony}, {Nandal}, {Sciarini}, {Hirai}, {Ekstr{\"o}m}, {Farrell}, {Murphy}, {Choplin}, {Hirschi}, {Chiappini}, {Liu}, {Bromm}, {Groh}, \& {Meynet}}]{Tsiatsiou2024}
{Tsiatsiou}, S., {Sibony}, Y., {Nandal}, D., {et~al.} 2024, \aap, 687, A307

\bibitem[{{Vink}(2018)}]{Vink2018}
{Vink}, J.~S. 2018, \aap, 615, A119

\bibitem[{{Vink}(2023)}]{Vink2023}
{Vink}, J.~S. 2023, \aap, 679, L9

\bibitem[{{Whalen} {et~al.}(2020){Whalen}, {Surace}, {Bernhardt}, {Zackrisson}, {Pacucci}, {Ziegler}, \& {Hirschmann}}]{Whalen2020}
{Whalen}, D.~J., {Surace}, M., {Bernhardt}, C., {et~al.} 2020, \apjl, 897, L16

\bibitem[{{Wise} {et~al.}(2019{\natexlab{a}}){Wise}, {Regan}, {O'Shea}, {Norman}, {Downes}, \& {Xu}}]{Wise2019}
{Wise}, J.~H., {Regan}, J.~A., {O'Shea}, B.~W., {et~al.} 2019{\natexlab{a}}, \nat, 566, 85

\bibitem[{{Wise} {et~al.}(2019{\natexlab{b}}){Wise}, {Regan}, {O'Shea}, {Norman}, {Downes}, \& {Xu}}]{Wise_2019}
{Wise}, J.~H., {Regan}, J.~A., {O'Shea}, B.~W., {et~al.} 2019{\natexlab{b}}, \nat, 566, 85

\bibitem[{{Woods} {et~al.}(2020){Woods}, {Heger}, \& {Haemmerl{\'e}}}]{Woods2020}
{Woods}, T.~E., {Heger}, A., \& {Haemmerl{\'e}}, L. 2020, \mnras, 494, 2236

\bibitem[{{Woods} {et~al.}(2017){Woods}, {Heger}, {Whalen}, {Haemmerl{\'e}}, \& {Klessen}}]{Woods2017}
{Woods}, T.~E., {Heger}, A., {Whalen}, D.~J., {Haemmerl{\'e}}, L., \& {Klessen}, R.~S. 2017, \apjl, 842, L6

\bibitem[{{Woosley} \& {Heger}(2006)}]{Woosley2006}
{Woosley}, S.~E. \& {Heger}, A. 2006, \apj, 637, 914

\bibitem[{{Wu} {et~al.}(2015){Wu}, {Wang}, {Fan}, {Yi}, {Zuo}, {Bian}, {Jiang}, {McGreer}, {Wang}, {Yang}, {Yang}, {Thompson}, \& {Beletsky}}]{Wu2015}
{Wu}, X.-B., {Wang}, F., {Fan}, X., {et~al.} 2015, \nat, 518, 512

\bibitem[{{Yoon} {et~al.}(2012){Yoon}, {Dierks}, \& {Langer}}]{Yoon2012}
{Yoon}, S.~C., {Dierks}, A., \& {Langer}, N. 2012, \aap, 542, A113

\bibitem[{{Yoon} \& {Langer}(2005)}]{Yoon2005}
{Yoon}, S.~C. \& {Langer}, N. 2005, \aap, 443, 643

\bibitem[{{Zahn}(1992)}]{Zahn1992}
{Zahn}, J.~P. 1992, \aap, 265, 115

\end{thebibliography}

\end{document}